\documentclass[fleqn,usenatbib,useAMS]{mnras}

\usepackage{threeparttable}

\usepackage{bm}
\usepackage{graphicx}	
\usepackage{amsmath}	
\usepackage{amssymb}	
\usepackage{multicol}        
\usepackage{bm}		
\usepackage{pdflscape}	
\usepackage{epsfig}

\usepackage[T1]{fontenc}
\usepackage{ae,aecompl}

\usepackage{newtxtext,newtxmath}

\title[Spectro-polarimetric analysis of GRB 160325A]{
Spectro-polarimetric analysis of prompt emission of GRB 160325A: jet with evolving environment of internal shocks
}
\author[V. Sharma et al.]{Vidushi Sharma,$^{1}$\thanks{E-mail: vidushi@iucaa.in} Shabnam Iyyani,$^{1}$\thanks{E-mail: shabnam@iucaa.in} Dipankar Bhattacharya,$^{1}$
Tanmoy Chattopadhyay,$^{2,3}$
\newauthor
Santosh V. Vadawale,$^{4}$ Varun. B. Bhalerao$^{5}$\\
\\
\normalsize{$^{1}$Inter-University Center for Astronomy and Astrophysics, Pune, Maharashtra 411007, India}\\
\normalsize{$^{2}$Department of Physics, Stanford University, 382 Via Pueblo Mall,
Stanford CA 94305}\\
\normalsize{$^{3}$Kavli Institute of Astrophysics and Cosmology, 452 Lomita Mall,
Stanford, CA 94305}\\
\normalsize{$^{4}$Physical Research Laboratory, Ahmedabad, Gujarat 380009, India}\\
\normalsize{$^{5}$Indian Institute of Technology Bombay, Mumbai, India}\\
}

\date{Accepted XXX. Received YYY; in original form ZZZ}

\pubyear{2019}
\begin{document}
\label{firstpage}
\pagerange{\pageref{firstpage}--\pageref{lastpage}}
\maketitle

\begin{abstract}
GRB 160325A is the only bright burst detected by {\it AstroSat} CZT Imager in its primary field of view to date. In this work, we present the spectral and polarimetric analysis of the prompt emission of the burst using {\it AstroSat}, {\it Fermi} and {\it Niel Gehrels Swift} observations.
The prompt emission consists of two distinct emission episodes separated by a few seconds of quiescent/ mild activity period.
The first emission episode shows a thermal component as well as a low polarisation fraction of $PF < 37\, \%$ at $1.5\, \sigma$ confidence level. On the other hand, the second emission episode shows a non-thermal spectrum and is found to be highly polarised with $PF > 43\, \%$ at $1.5 \sigma$ confidence level.
We also study the afterglow properties of the jet using {\it Swift}/XRT data. The observed jet break suggests that the jet is pointed towards the observer and has an opening angle of $1.2^{\circ}$ for an assumed redshift, $z = 2$. With composite modelling of polarisation, spectrum of the prompt emission and the afterglow, we infer that the first episode of emission originates from the photosphere with localised dissipation happening below it, and the second from the optically thin region above the photosphere. The photospheric emission is generated mainly by inverse Compton scattering, whereas the emission in the optically thin region is produced by the synchrotron process. The low radiation efficiency of the burst suggests that the outflow remains baryonic dominated throughout the burst duration with only a subdominant Poynting flux component, and the kinetic energy of the jet is likely dissipated via internal shocks which evolves from an optically thick to optically thin environment within the jet.
\end{abstract}

\begin{keywords}
gamma-ray burst: individual (GRB 160325A) -- {\it AstroSat} -- {\it Fermi} -- {\it Swift} -- Spectral analysis -- polarisation
\end{keywords}

\section{Introduction}
Gamma-ray bursts (GRBs) are the most energetic transient events in the universe followed by the formation of compact objects. The initial brief and intense flash of gamma rays, known as prompt emission, originates close to the burst site. The most popular model to explain the GRB dynamics is the relativistic fireball model (\citealt{Paczynski_1986,Piran_1993}, for a detailed review of the model please refer \citealt{meszaros2006gamma,peer2015,Kumar_Zhang2015}), where the high temperature fireball formed after the explosion expands due to high radiation pressure and thereby accelerates the outflow to a saturation radius where the internal energy density becomes equal to the kinetic energy density of the outflow. At some point, the optical depth of the outflow becomes equal to unity which is referred to as the photosphere, where the photons get decoupled from the plasma. A thermalized emission is expected from the photosphere and later on in the optically thin region above the photosphere, the kinetic energy of the jet gets dissipated via internal shocks constituting the dominant part of the observed gamma ray emission \citep{rees1994unsteady,Daigne_Mochkovitch1998}. The outflow finally then runs into the external ambient medium producing afterglow emission as the outflow dissipates its remaining energy through external shocks \citep{paczynski1993radio,piran1998spectra}.

With the launch of sensitive space based detectors like Burst And Transient Source Experiment (BATSE, \citealt{fishman1989batse}), {\it Fermi} \citep{meegan2009fermi,atwood2009large} and {\it Niel Gehrels Swift} \citep{gehrels2004swift}, the knowledge of spectral properties of the prompt emission of GRBs has improved significantly. However, the nature of the emission mechanism still remains inconclusive. The non-thermal nature of the GRB spectra are generally attributed to radiation processes such as synchrotron \citep{rees1994unsteady} or inverse Compton scattering \citep{ghirlanda2003extremely}. However, these non-thermal models alone are unable to explain certain spectral features like hard low energy power law index \citep{gruber2014fermi}, narrow peak energy distribution \citep{gruber2014fermi} and also the high radiation efficiency observed in bright bursts \citep{kobayashi1997can}. Photospheric emission, which is inherently present in the fireball model, was thus invoked to resolve some of these issues. 
Studies conducted by \citealt{ryde2005thermal,ryde2009quasi}, \citealt{ryde2010identification,guiriec2011detection,axelsson2012grb110721a}, 
\citealt{iyyani2013variable}, 
\citealt{burgess_etal_2014a},
\citealt{iyyani2015extremely} 
and \citealt{Acuner_etal_2019} 
found that many GRBs were best modelled using a combination of thermal (blackbody function) and non-thermal (power law, Band function \citep{band1993batse} or synchrotron) spectral functions. These detections also supported the idea that within the picture of classical fireball model, the thermal emission from the photosphere can be either subdominant or sometimes dominant depending on how much adiabatic cooling the outflow has undergone in the coasting phase before it reaches the photosphere. 

The spectral analysis of GRB 090902B showed that a narrow hard spectrum evolved into a broader one with time \citep{ryde2010identification,Ryde_etal_2011,Omer_etal_2011,Begue_Iyyani2014}. This pointed towards the possibility of subphotospheric dissipation, which when functional throughout the outflow till the photosphere, results in a significant broadening of the spectrum from that of a thermal emission and can resemble a typical Band function \citep{Giannios2008,Lazzati_etal_2009,Beloborodov2011,beloborodov2013regulation,Begue_Pe'er2015}. On the other hand, it was shown in the case of GRB 110920A that when a localized subphotospheric dissipation happens at a lower optical depth (closer to the photosphere), the observed spectrum from the photopshere would not look smooth like a Band function but can instead possess multiple spectral breaks resulting in a top-hat like shape \citep{Pe'er_Waxman2004,Pe'er_Waxman2005,iyyani2015extremely}. In both these bursts, along with the dominant photospheric emission, a non-thermal component (power law) was found throughout the burst duration which was related to non-thermal emission produced in the optically thin region. Thus, in several previous studies different spectral shapes and components have been related to emissions coming from different regions in the outflow. 

Linear polarisation measurement of the prompt emission of GRBs is a powerful tool, in addition to spectral analysis, to strongly constrain the radiation process, geometry of the emitting region and the composition of the GRB outflow \citep{covino2016polarization}. Currently, polarisation measurement of prompt emission has been reported for only a handful of cases (\citealt{wigger2004gamma,willis2005evidence,mcglynn2007polarisation,gotz2009variable,mcglynn2009high, yonetoku2011detection,yonetoku2012magnetic,gotz2013polarized, gotz2014grb, zhang2019detailed,burgess_etal_2019,Chand2018,Chand2019,2019ApJ_sharma,2019ApJ_Chattopadhyay}; for a recent review see \citealt{mcconnell2017high}). Linear polarisation has also been measured in early optical afterglows of a few GRBs \citep{Wijers_1999_990510_afterglowpol,Covino_2002_011211,Greiner_2003_030329_opticalpol,King2014_131030A_opticalpol,Gorbovskoy_2016_opticalpol,troja2017significant,Laskar_etal_2019}.
Polarised emission is generally associated with some asymmetry of the emission within the emitting or the viewing cone of the jet \citep{Toma_etal_2009}.

The Cadmium Zinc Telluride Imager (CZTI; \citealt{vadawale2015hard}) on board {\it AstroSat} \citep{singh2014astrosat} is a new addition to the existing instruments studying GRB science. CZTI utilizes coded aperture mask for localising sources within its primary $4^{\circ}.6 \times 4^{\circ}.6$ field of view and is spectroscopically verified for on-axis sources in $20-200$ keV range (\citealt{rao2016astrosat}, \citealt{bhalerao2017cadmium}). At energies above $100\, \rm keV$, CZTI behaves like an open detector with all sky response.  
It also provides a unique opportunity of studying hard X-ray polarisation in $100-400$ keV energy range, due to its capability of identifying Compton scattered events in the case of bright bursts. The selection procedure for finding the Compton events (a dominant mechanism for photon detection in this energy range) in CZTI is described in detail in \citealt{chattopadhyay2014prospects}. In the first year of CZTI operation, polarisation measurements were performed for 11 bright GRBs which had sufficient number of such Compton scattered events \citep{2019ApJ_Chattopadhyay}.

In this paper, we present the spectro-polarimetric analysis and physical modelling of the long bright GRB 160325A, which is the only gamma-ray burst detected in the field of view of {\it AstroSat}/CZTI to date and also concurrently detected by
{\it Swift} Burst Alert Telescope (BAT) and {\it Fermi} Gamma-ray Space Telescope with its Gamma-ray Burst Monitor (GBM).
The paper is arranged in the following manner: In Section 2, we present the observations and main features of the burst. We present the joint time-resolved spectral analysis of the prompt-emission phase in Section 3 using BAT and GBM data, followed by the joint spectral and polarimetry analyses for the two episodes of GRB 160325A using CZTI data in Section 4. The {\it Swift} X-Ray Telescope (XRT) afterglow observations are presented in Section 5. We then finally discuss our results and physical modelling in section 6 and summarise the observations and our conclusions in Section 7.

We have used $\Lambda$CDM cosmology model with cosmological parameters, $H_{0} = 67.4 \pm 0.5$ km/s/Mpc, $\Lambda_{vac} = 0.685$ and $ \Lambda_{m} = 0.315$  (Planck collaboration 2018, \citealt{2018_planck}) for calculating the energetics of the GRB. Since there was no redshift available for GRB 160325A, we have assumed a redshift of $z = 2$, which is roughly the average redshift of long GRBs \citep{Bagoly_etal_2006,Wang_etal_2019,Lloyd-Ronning_etal_2019}.

\section{Observations and Lightcurve}
GRB 160325A was discovered and localized by {\it Niel Gehrels Swift}/BAT on 2016 March 25 at 07:00:03 UT \citep{Swift_BAT_GCN_160325}. {\it Swift}/XRT started observing the field at nearly $66$ s after the BAT trigger and found a bright, fading uncatalogued X-ray source. {\it Swift} UVOT also made optical observations for this burst from as early as $75\, \rm s$ post BAT trigger, signifying the early afterglow emission \citep{2016GCN_UVOT}. Using XRT-UVOT alignment, the astrometrically corrected X-ray position of the burst was RA, Dec = 15.65055, -72.69659 with an uncertainty of 1.7 arcsec \citep{Swift_XRT_GCN_160325}. At 06:59:21.51 UT ($T_0$) {\it Fermi} Gamma-Ray Burst Monitor (GBM) also triggered and located this burst \citep{Fermi_GBM_GCN_160325}. The {\it Fermi} Large Area Telescope (LAT) detected high energy emission ($>100 \, \rm MeV$) at nearly $5$ s post trigger ($T_0$) 
\citep{Fermi_LAT_GCN_160325}. The burst was also detected by {\it AstroSat} Cadmium Zinc Telluride Imager (CZTI) at 06:59:21.51 UT and $Konus-Wind$ \citep{Konus_GCN_160325}. GRB 160325A was the only GRB observed in the primary field of view (FOV) of the CZTI on-board AstroSat in its first three years of operation. This burst was incident at $\theta = 0.66^{\circ}$ and $\phi = 159.45^{\circ}$ from the pointing direction of the CZTI.

\begin{figure}
	\includegraphics[width=\columnwidth]{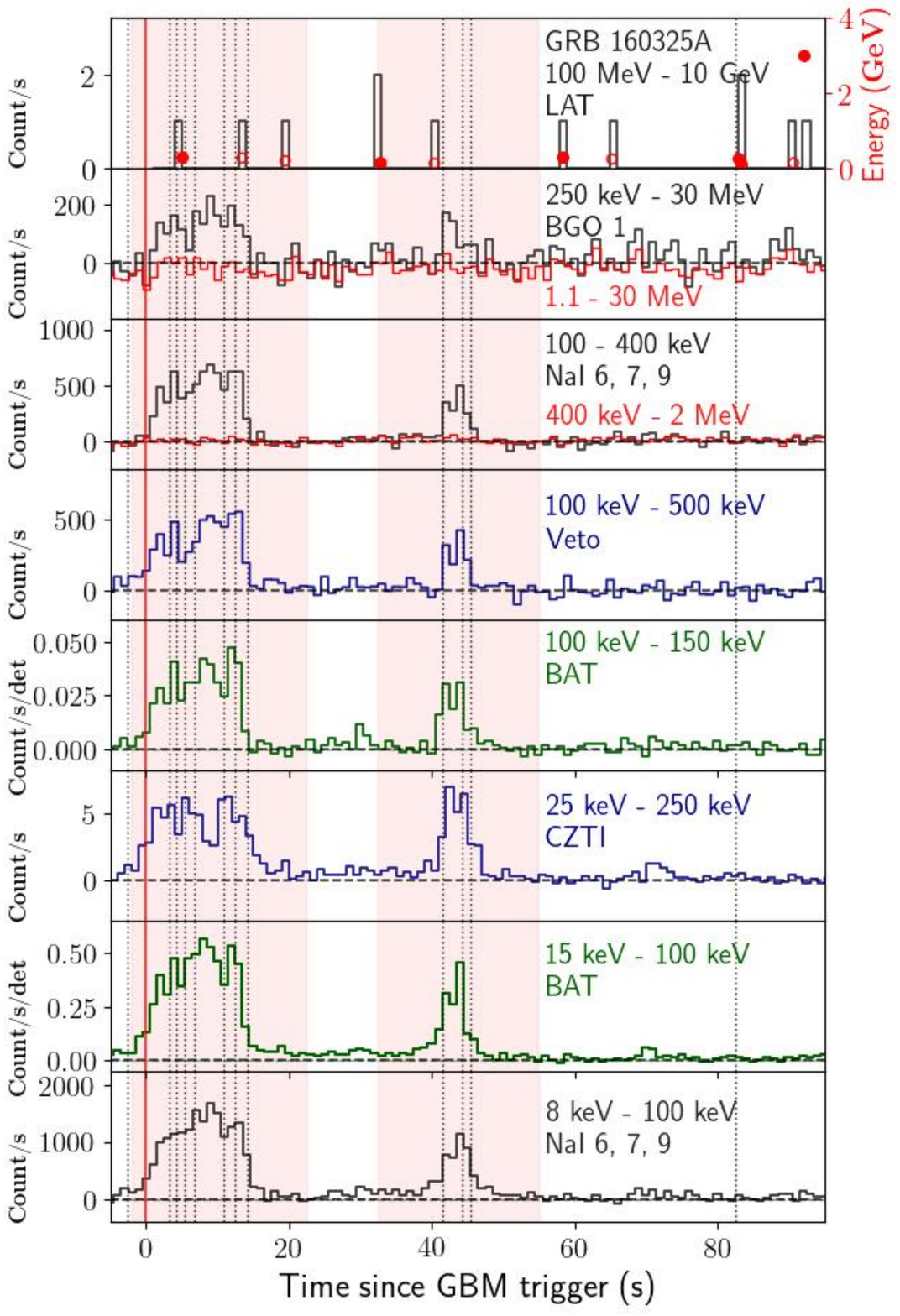}
    \caption{A composite 1 second binned lightcurve of the burst is shown. This includes {\it Fermi} detectors: LAT $\&$ GBM (NaI 6, 7, 9 $\&$ BGO 1); {\it AstroSat} detectors: CZTI $\&$ Veto and {\it Swift}/BAT. The red vertical solid line represents the GBM trigger time and black vertical dotted lines correspond to the time ranges used for spectral analysis. The {\it Fermi}/LAT detected photons are shown as red points in the upper most panel, where the energy information of each photon is shown on y-axis located at the right side of the panel. The LAT detected photons having a probability of association with the same GRB to be greater and less than $68\%$ are shown in red solid points and red hollow points, respectively.}
    \label{fig:figure_lc}
\end{figure}

A composite lightcurve of the prompt emission of the GRB is shown in Figure~\ref{fig:figure_lc} which includes data from {\it Fermi}, {\it AstroSat} and {\it Swift}. Bottom to top panels in the plot correspond to the photon count rate in different energy ranges spanning from $8$ keV to a few GeV. The lightcurve consists of two emission episodes separated by a period of $\sim 9$ s 
of mild soft emission dominantly in the energy range $<100 \, \rm keV$. 
It is, thus, unclear as to when exactly the first emission episode ends and the second emission episode starts. By studying the light curves observed in the different detectors, we have defined the first and second emission episodes as the emission received during the intervals  $-2 - 23 \, \rm s$ and $32 - 55 \, \rm s$ respectively relative to the {\it Fermi} trigger reference. The time duration ($T_{90}$) of the burst is found to be $43 \pm 0.6$ ($64.9 \pm 14.7$) s by {\it Fermi}/GBM ({\it Swift}/BAT), which is the time taken to accumulate $90\%$ of the total fluence of the burst. 
The background subtracted count rates are shown for all detectors in the lightcurve.
We also note that there is no significant emission detected above $1.1$ MeV (second panel from above in Figure \ref{fig:figure_lc}). 
The signal in the energy range of $400$ keV to $2$ MeV and $1.1$ to $30$ MeV (plotted in solid red lines) seen by NaI detectors and BGO 1 detector respectively is found to be consistent with background. Thus, NaI and BGO 1 data within energy range 8 - 400 keV and 300 keV - 4 MeV respectively were used for spectral analyses. 

\section{Spectral Analysis of the prompt emission}
Time-resolved spectral analyses using {\it Swift}/BAT and {\it Fermi}/GBM data were performed.
We defined the time intervals for the analyses using Bayesian Blocks algorithm \citep{scargle1998studies,Burgess2014} and obtained 11 blocks of different count rates for the {\it Swift}/BAT detector in which the count rate was the largest. This task was performed using the {\it Swift} battblocks tool \footnote{https://heasarc.gsfc.nasa.gov/ftools/caldb/help/battblocks.html}.
We used standard BAT procedures\footnote{https://swift.gsfc.nasa.gov/analysis/threads/bat$\_$threads.html} to extract the lightcurves and background-subtracted spectra from the event file.

For the analysis of {\it Fermi}/GBM observations, three NaI detectors with high count rates were chosen. The NaI detectors are denoted by nx where x refers to the number of the NaI detector. For this burst NaI detectors, namely n6, n7 and n9 had smaller source angles $<60^{\circ}$ \citep{gruber2014fermi} and hence were chosen for analysis. Since the number of all NaI detectors (nx) were x > 5, we chose the BGO\footnote{https://fermi.gsfc.nasa.gov/ssc/data/analysis/scitools/rmfit$\_$tutorial.html$\#$GBM}, b1. LAT data ($> 100 \, \rm MeV$) were not used in the analysis as there were not enough number of photons whose probability of association with the GRB was greater than $90\%$. Using {\it Fermi} Burst Analysis GUI v 02-01-00p1 (gtburst3 \footnote{https://fermi.gsfc.nasa.gov/ssc/data/analysis/scitools/gtburst.html}), spectra for each {\it Fermi} detector were made. 
The background was estimated from the regions before and after the GRB for the time intervals $T_{0}$ - 125 s to $T_{0}$ - 10 s and $T_{0}$ + 85 s to $T_{0}$ + 200 s, respectively. 
Joint spectral analyses of BAT and {\it Fermi} data were done using X-Ray Spectral Fitting Package (XSPEC) \citep{arnaud1996xspec} version: 12.9.0n. 
As the obtained spectral files for BAT follow the Gaussian statistics, whereas the spectral files for GBM are consistent with statistics for Poisson data with Gaussian background (pgstat), we rebinned the GBM spectra using Heasoft tool grppha\footnote{https://heasarc.gsfc.nasa.gov/ftools/caldb/help/grppha.txt} \citep{virgili2012spectral} with the criterion that each bin consists of at least 20 photon counts.
Thus, $\chi^{2}$ - statistics was used for the joint spectral analyses.

\subsection{Effective Area Correction}
To estimate the effective area correction between the different detectors of {\it Fermi} and BAT, 
we adopted the following methodology \citep{page2009multiwavelength}: (a) To find the best fit model: Each of the time resolved spectra of all detectors (BAT, GBM: n6, n7, n9 and b1) was jointly analysed using different spectral functions such as power law, cutoff-power law and Band, without including the constant factor correction. In this exercise, the time-averaged and time-resolved spectra were found to be better fitted by cutoff-powerlaw function. (b) Finding the normalization-offsets: We then simultaneously fitted the five bright time resolved spectra with the best fit model, cutoff-powerlaw, by multiplying the model by a constant of normalization for each instrument. The constant of BAT data was frozen to unity, while those of the GBM detectors were kept free. In addition, these constants were also tied across the respective data sets of the different time-resolved intervals. 
The relative normalizations were found as, $1.14^{+0.03}_{-0.03}$ for n6 detector, $1.09^{+0.03}_{-0.03}$ for n7 detector, $1.07^{+0.03}_{-0.03}$ for n9 detector and $1.09^{+0.17}_{-0.16}$ for b1 detector. These constant factors are consistent with the expected inter-calibration uncertainty of up to $30\%$ and hence were used for the subsequent spectral analyses.

\subsection{Parameters Evolution}
\begin{figure}
	\includegraphics[width=\columnwidth]{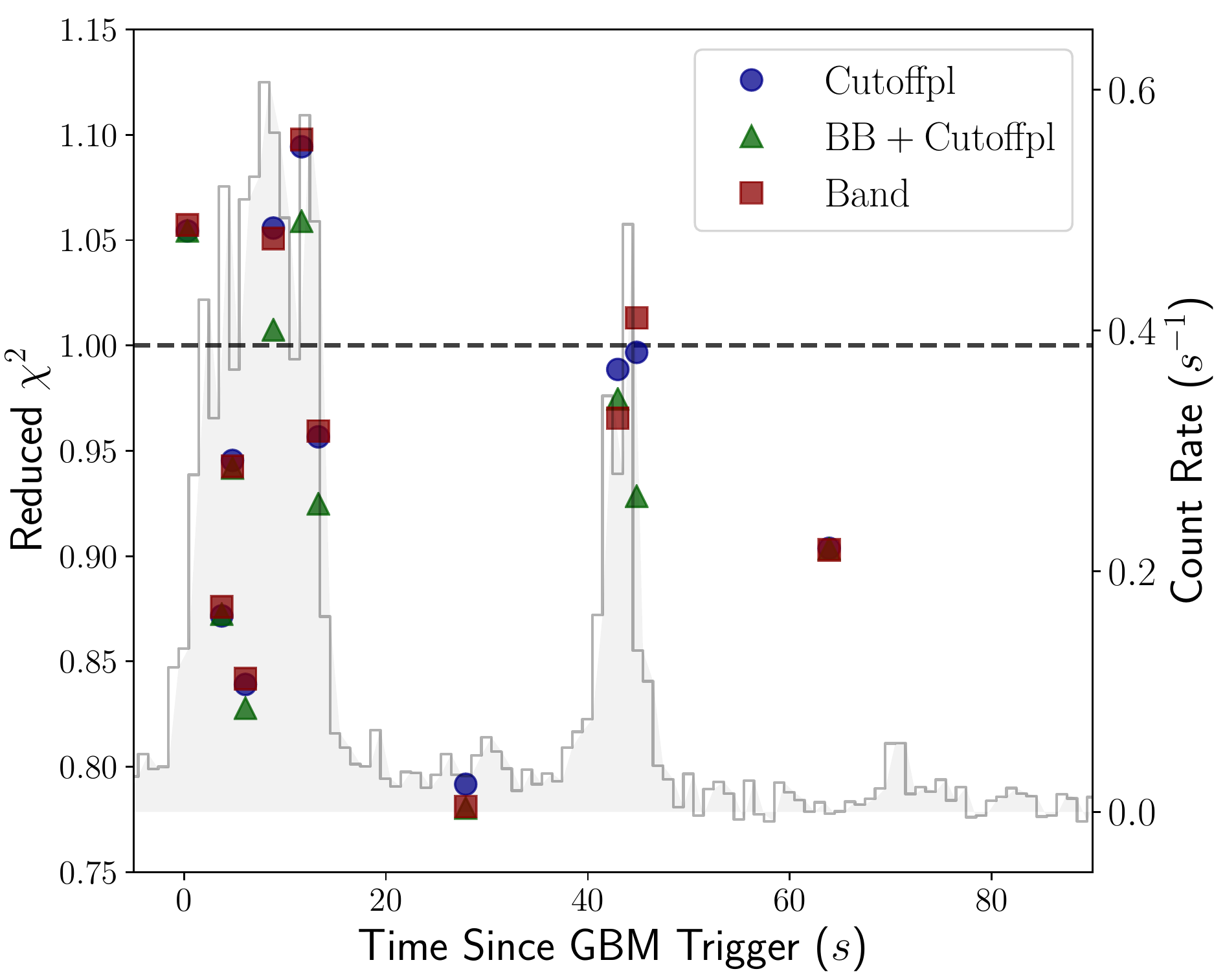}
    \caption{The reduced $\chi^{2}$-statistics obtained for the three spectral models are shown. The blue circles present the cutoff-powerlaw, while the green triangle is for the blackbody+cutoff-powerlaw model and Band model is shown by red squares. In the background in grey colour the {\it Swift}/BAT lightcurve obtained in 15-150 keV energy range is shown.}
    \label{fig:figure_red_chi}
\end{figure}

For characterising the spectra, we fit the brightest time intervals in
both episodes with several phenomenological models including power law
with an exponential cutoff (cutoff-power law), blackbody + cutoff-power
law, and traditional Band functions.
In Figure~\ref{fig:figure_red_chi}, the reduced $\chi^{2}$ obtained for the different models in the different time intervals are shown.
We note that in the brightest time interval (bin $\#$ 5) of the first episode, blackbody + cutoff-powerlaw model is statistically favoured (i.e reduced $\chi^2 \sim 1$). On the addition of blackbody component to the cutoff-power law, we also find an improvement in Akaike Information Criterion (AIC; \citealt{akaike1974}) and Bayesian Information Criterion (BIC; \citealt{schwarz1978}) by 15 and 7 respectively (see Appendix B). At the same time, in all other time intervals, this model is also found to give comparable statistics when compared with other models. The Monte Carlo (MC) simulations for estimating the statistical significance of the additional blackbody component with the cutoff-powerlaw model are performed for the brightest time interval. The difference of $\chi^{2}$-statistics between cutoff-powerlaw model and blackbody + cutoff-powerlaw model is found to be $19$. Here, we assume cutoff-powerlaw model as the null hypothesis model and blackbody + cutoff-powerlaw model as the alternate model and simulations are run for $>15000$ realizations. We find that the probability of observing $\Delta \; \chi^{2}$ > $19$ is $10^{-2.52}$, which corresponds to $\sim 3\; \sigma$ significance level.
Thus, by considering the above fit statistics and MC simulations obtained in the brightest time bin, we consider blackbody + cutoff-power law model to be the best fit model for all the time resolved intervals of the first episode.

\begin{figure}
	\includegraphics[width=\columnwidth]{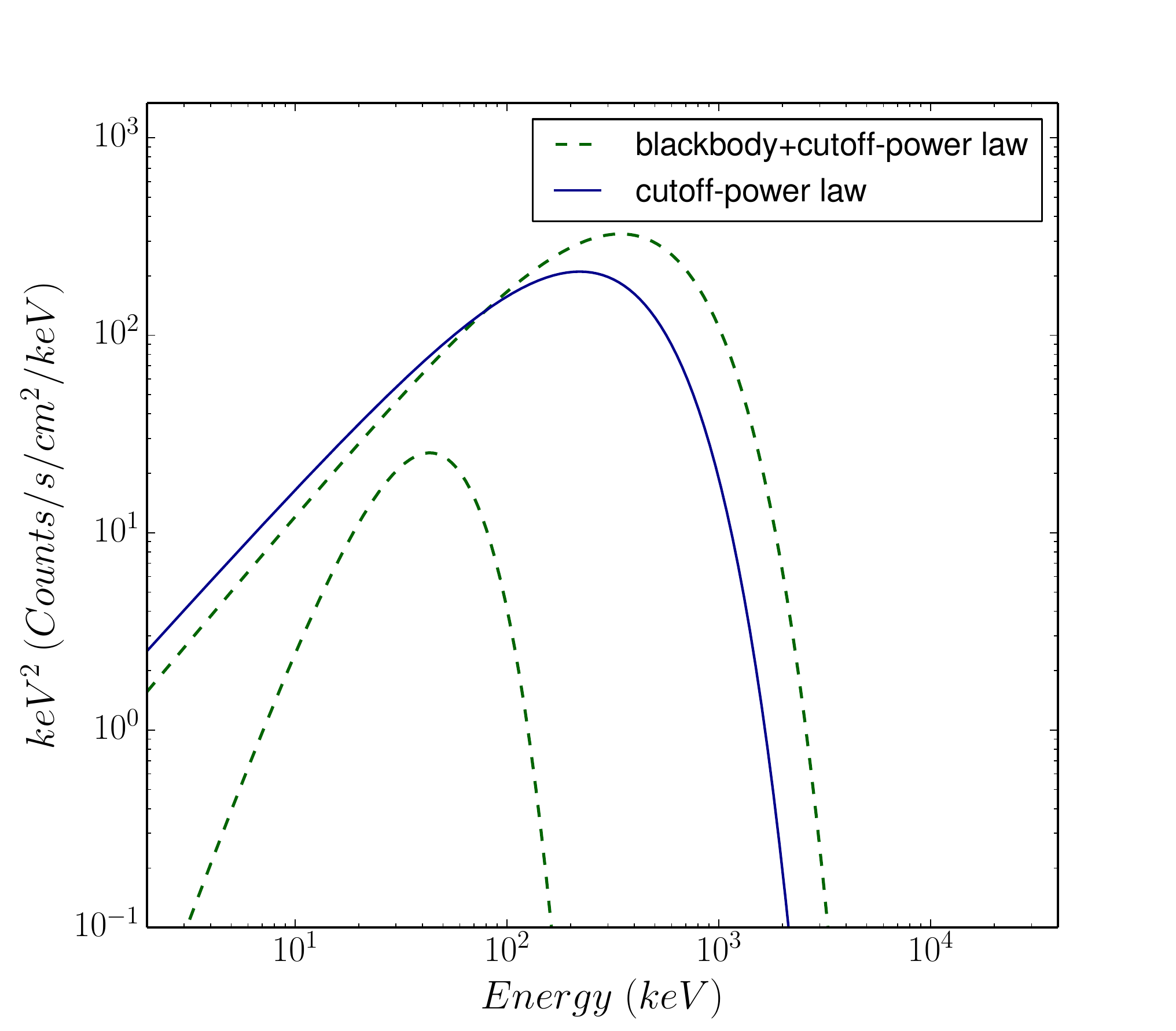}
    \caption{The $\nu F_{\nu}$ plots of a blackbody+cutoff-powerlaw model for the bright bin of the first episode and cutoff-powerlaw model for the second episode bright bin is shown in green dotted and blue solid lines, respectively.}
    \label{fig:spectral_model}
\end{figure}

\begin{figure*}
\centering
\begin{minipage}{.5\textwidth}
  \centering
  \includegraphics[width=1.0\linewidth]{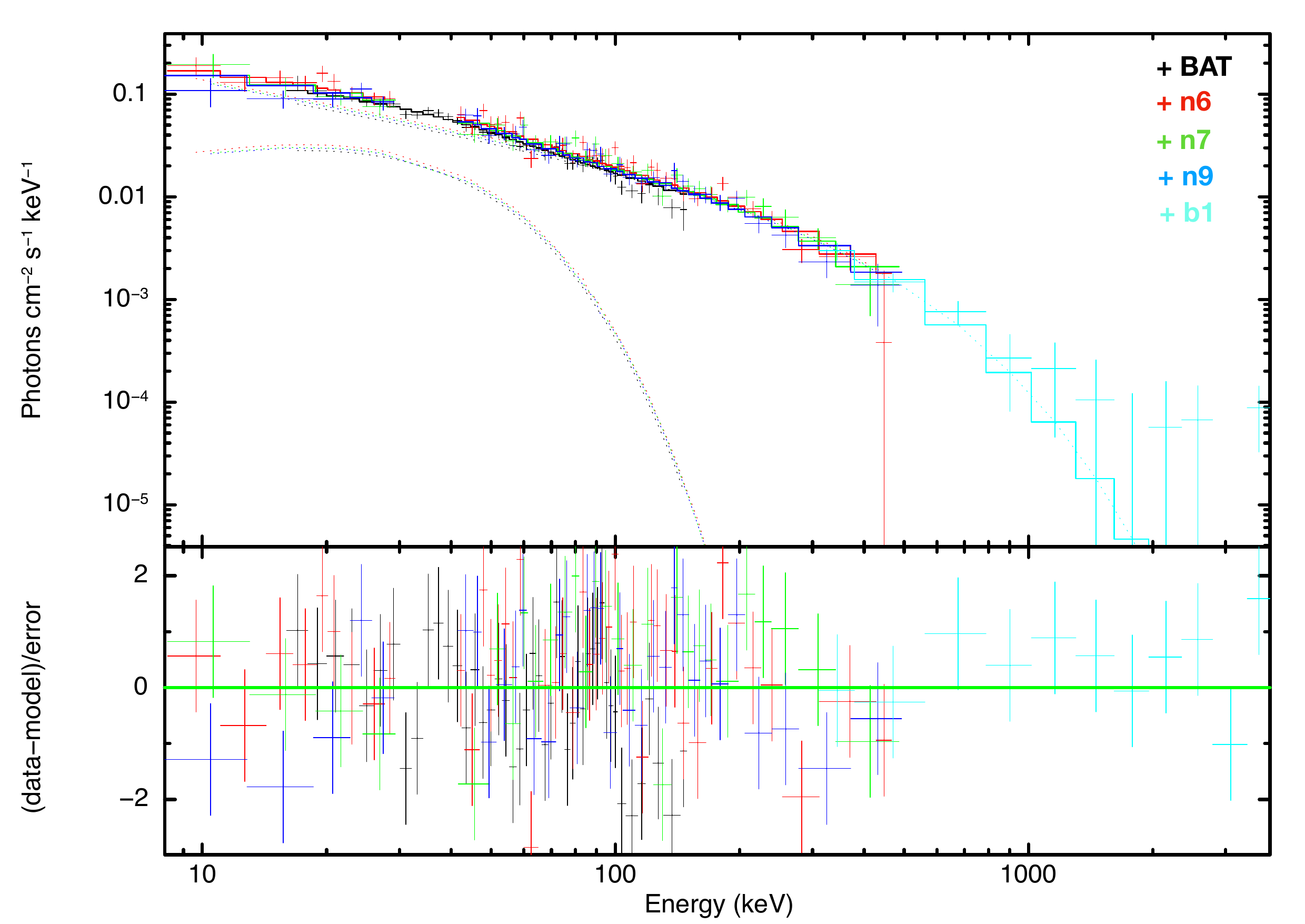}
\end{minipage}%
\begin{minipage}{.5\textwidth}
  \centering
  \includegraphics[width=1.0\linewidth]{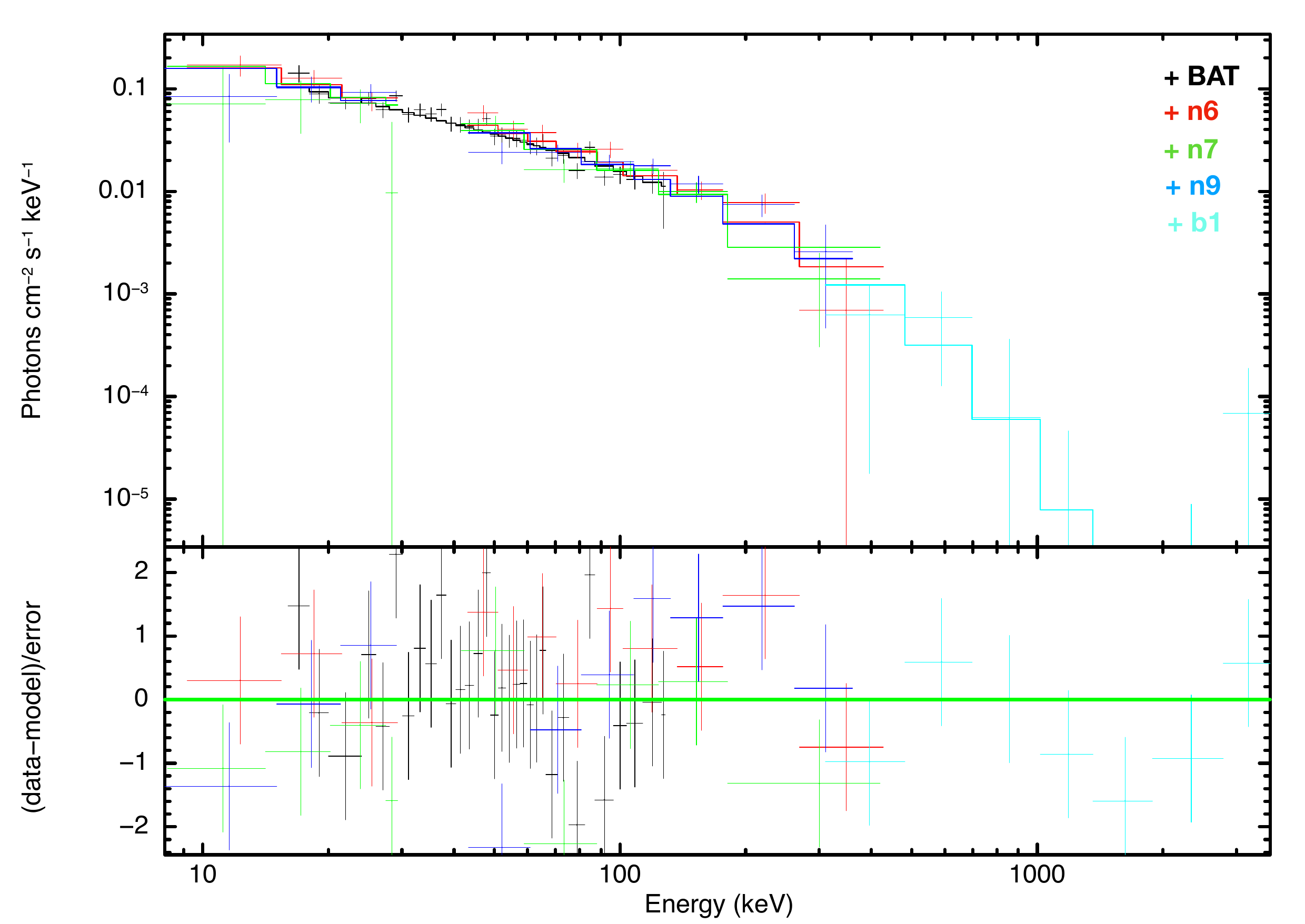}
\end{minipage}
\caption{Counts spectra for the brightest time bin of the first and second episodes are shown for the models blackbody + cutoff-powerlaw (left) and the cutoff-power law (right). The respective residuals obtained for the fit is shown in the lower panels of both the plots. 
}
\label{fig:160821A_spectral_fit}
\end{figure*}

During the second episode, the brightest interval is best modelled using a 
cutoff-power law model only, by giving a reduced $\chi^{2}$ closest to 1. 
On the addition of a blackbody component to cutoff-power law, the AIC shows an improvement by 1 and BIC shows an increment by 7, which suggests a less parameter model to be better (see Appendix B). 
Based on the above obtained fit statistics, the best fit model of the second emission episode is chosen to be cutoff-power law. The $\nu F_{\nu}$ plots of the best fit
model for the brightest time interval of the first and second emission episodes are shown in Figure ~\ref{fig:spectral_model}. The counts spectra for the bright bins of both the episodes and the residuals obtained from the fits are shown in Figure ~\ref{fig:160821A_spectral_fit}. We have reported the fit results of blackbody + cutoff-powerlaw and cutoff-powerlaw models in Table~\ref{tab:table1_cutoffpowerlaw_bb} and Table~\ref{tab:table1_cutoffpowerlaw}, respectively. 

\begin{figure}
	\includegraphics[width=\columnwidth]{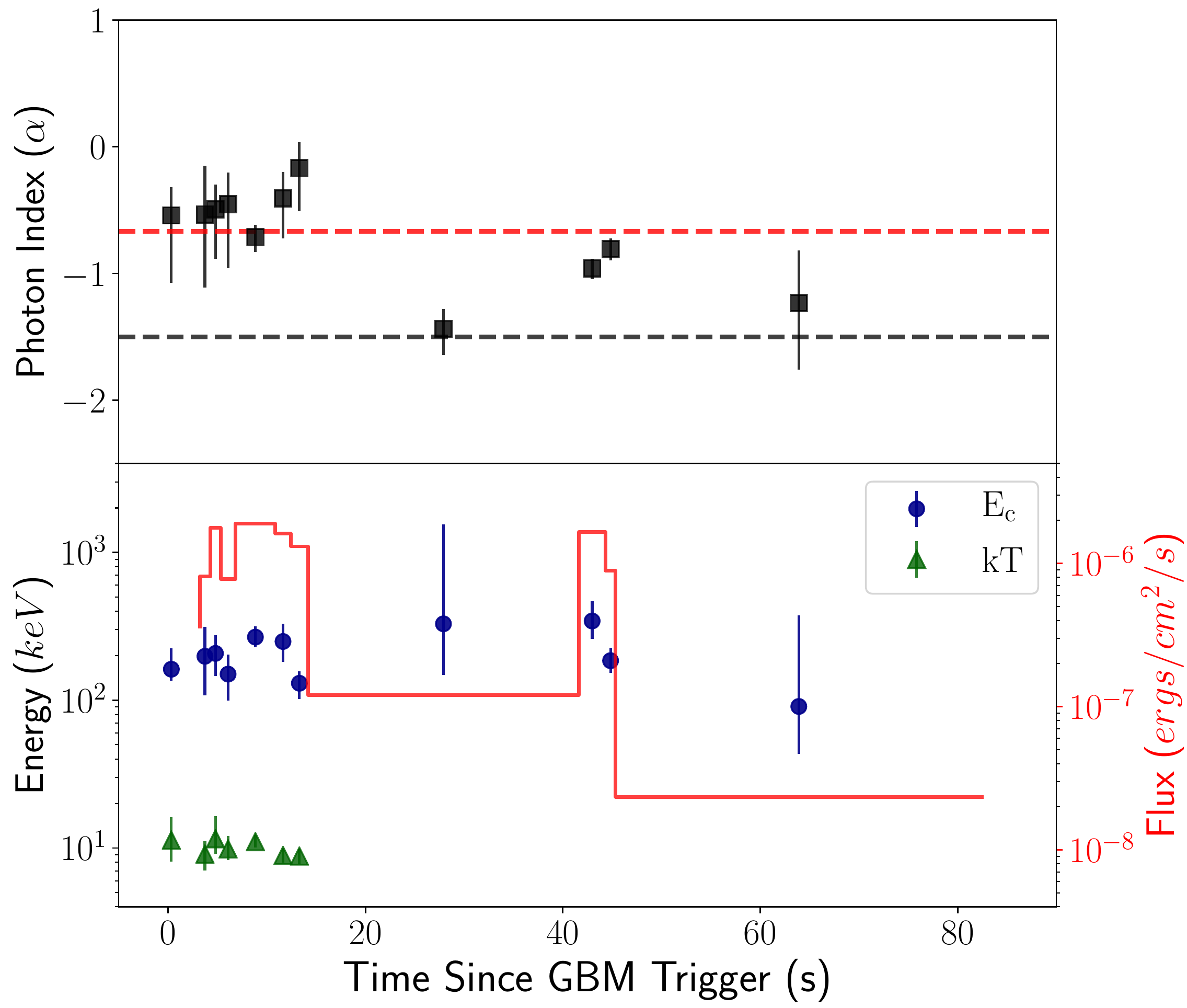}
    \caption{Temporal evolution of the parameters of the spectral models: blackbody+cutoff-powerlaw during the first episode and cutoff-powerlaw during the second episode are shown.
    In the upper panel, photon index ($\alpha$, black squares) is plotted along with the synchrotron fast cooling photon index of $-3/2$ (black dashed line) and the line of death of synchrotron emission i.e., $-2/3$ (red dashed line). In the lower panel, cutoff energy ($E_{c}$) and blackbody temperature ($kT$) are shown in blue circles and green triangles, respectively. Energy flux evolution is shown in solid red line and is scaled on right side of y-axis.}
    \label{fig:bb_cutoff_to_cutoff}
\end{figure}

 In Figure~\ref{fig:bb_cutoff_to_cutoff}, we have shown the temporal evolution of the parameters of the best-fit models. 
 With the classical fireball model, the non-thermal part of the spectrum is expected to be produced via process like synchrotron emission in the optically thin region. Thus, after adding a blackbody function to the spectrum, the power law index of the non-thermal component (here it is cutoff power law) is expected to get softer with time (\citealt{guiriec2013evidence}, also see \citealt{Burgess_etal_2015}).
 However, here we observe that the photon index ($\alpha$, upper panel of Figure~\ref{fig:bb_cutoff_to_cutoff}),is found to increase with time during the first episode and mostly lie above the line of death of standard model of synchrotron emission  (\citealt{Rybicki_Lightman1986,sari1998, preece1998synchrotron}, also see \citealt{burgess_etal_2019NatAs}) i.e. $\alpha = -2/3$. The thermal component shows hints of cooling from about $10 \, \rm keV$ to $8 \, \rm keV$. The ratio of thermal flux ($F_{T}$) to non-thermal flux ($F_{NT}$) in the observed energy range (8 keV - 1 MeV) is found to be $2\%-10\%$.
 During the second episode, the photon index is relatively much softer and is consistent with standard model of slow cooling synchrotron emission (also see \citealt{Uhm_zhang2015,burgess_etal_2019NatAs}). The cutoff energy shows an intensity tracking behaviour and follows the flux lightcurve. In prompt phase, the energy fluence which is the energy flux integrated over the time interval $T_{0}-2.54\, s$ to $T_{0}+82.46 \, s$, is estimated to be $1.90\times10^{-5} \; \mathrm{erg/cm^2}$ in $10-1000$ keV. 
 
\begin{table*}
\begin{small}
\caption{Parameters of time resolved spectral analysis of the best fit model of blackbody + Cutoff powerlaw in the first emission episode. }
\label{tab:table1_cutoffpowerlaw_bb}
\begin{center}
\begin{tabular}{p{1.25cm}ccccc}
\hline 
Bin $\#$	& Time Interval (s) & $kT$ (keV)	& - $\alpha$		& $E_{c}$ (keV)		& Reduced $\chi^{2}$-statistics                                          
\\ \hline \hline
1	& -2.54 - 3.21	& $11.27_{-3.16}^{+4.82}$  & $0.54_{-0.53}^{+0.22}$	& $162_{-27}^{+63}$	& 1.05
\\ 
2	& 3.21 - 4.27	& $9.07_{-2.00}^{+2.06}$  & $0.53_{-0.58}^{+0.38}$	& $198_{-90}^{+115}$	& 0.87
\\
3	& 4.27 - 5.34	& $11.53_{-2.42}^{+4.93}$  & $0.49_{-0.39}^{+0.19}$	& $207_{-61}^{+67}$	& 0.94
\\ 
4	& 5.34 - 6.83	& $9.84_{-1.56}^{+2.19}$  & $0.45_{-0.50}^{+0.24}$	& $150_{-51}^{+53}$	& 0.83
\\ 
5	& 6.83 - 10.87	& $11.03_{-0.93}^{+1.22}$  & $0.71_{-0.12}^{+0.10}$	& $267_{-39}^{+48}$	& 1.01
\\ 
6	& 10.87 - 12.42	& $8.90_{-0.94}^{+0.95}$  & $0.41_{-0.31}^{+0.21}$	& $250_{-68}^{+80}$	& 1.06
\\
7	& 12.42 - 14.19	& $8.81_{-1.03}^{+1.05}$  & $0.17_{-0.34}^{+0.20}$	& $130_{-28}^{+26}$	& 0.92
\\ \hline \hline
\end{tabular}
\end{center}
\end{small}
\end{table*}

\begin{table*}
\begin{small}
\caption{Parameters of time resolved spectral analysis of the best fit model of cutoff powerlaw in second emission episode.}
\label{tab:table1_cutoffpowerlaw}
\begin{center}
\begin{tabular}{p{1.25cm}ccccc}
\hline \hline
Bin $\#$	& Time Interval (s)		& - $\alpha$		& $E_{c}$ (keV)		& Reduced $\chi^{2}$-statistics                                          
\\ \hline 
8	& 14.19 - 41.62	& $1.44_{-0.21}^{+0.16}$	& $329_{-180}^{+1225}$	&  0.79
\\ 
9	& 41.62 - 44.32	& $0.96_{-0.09}^{+0.08}$	& $344_{-83}^{+122}$	& 0.99
\\ 
10	& 44.32 - 45.35	& $0.81_{-0.09}^{+0.08}$	& $185_{-32}^{+41}$	& 0.99
\\
11	& 45.35 - 82.46	& $1.23_{-0.53}^{+0.41}$	& $907_{-47}^{+283}$	& 0.90
\\ \hline \hline
\end{tabular}
\end{center}
\end{small}
\end{table*}
\section{Spectroscopy and polarisation study with \textit{AstroSat}/CZTI}
GRB 160325A is viewed within the field of view of CZTI. 
The on-axis effective area of CZTI is
significantly higher than off-axis, and the instrument response is very
well characterised.  
Enabled by this, here we have used CZTI data for performing the spectro-polarimetric analysis of the time integrated data of each emission episode of the GRB along with the data from $\it Fermi$ and $\it Swift$.

\subsection{Joint Spectral Analysis}
The CZTI detector onboard {\it AstroSat} utilises a coded aperture mask telescope and Cadmium Zinc Telluride detectors in $20-200$ keV energy range \citep{bhalerao2017cadmium}. CZTI also functions as an open detector by being sensitive to almost all sky (except Earth occulted region) for energies $>$ 100 keV. The time-tagged photon information like the position of photon interaction (pixel ID), energy and time of the registered photons, etc. are recorded at CZT detectors. It consists of four quadrants, and data for each quadrant is available separately. 

\begin{figure*}
\centering
\begin{minipage}{.5\textwidth}
  \centering
  \includegraphics[width=1.0\linewidth]{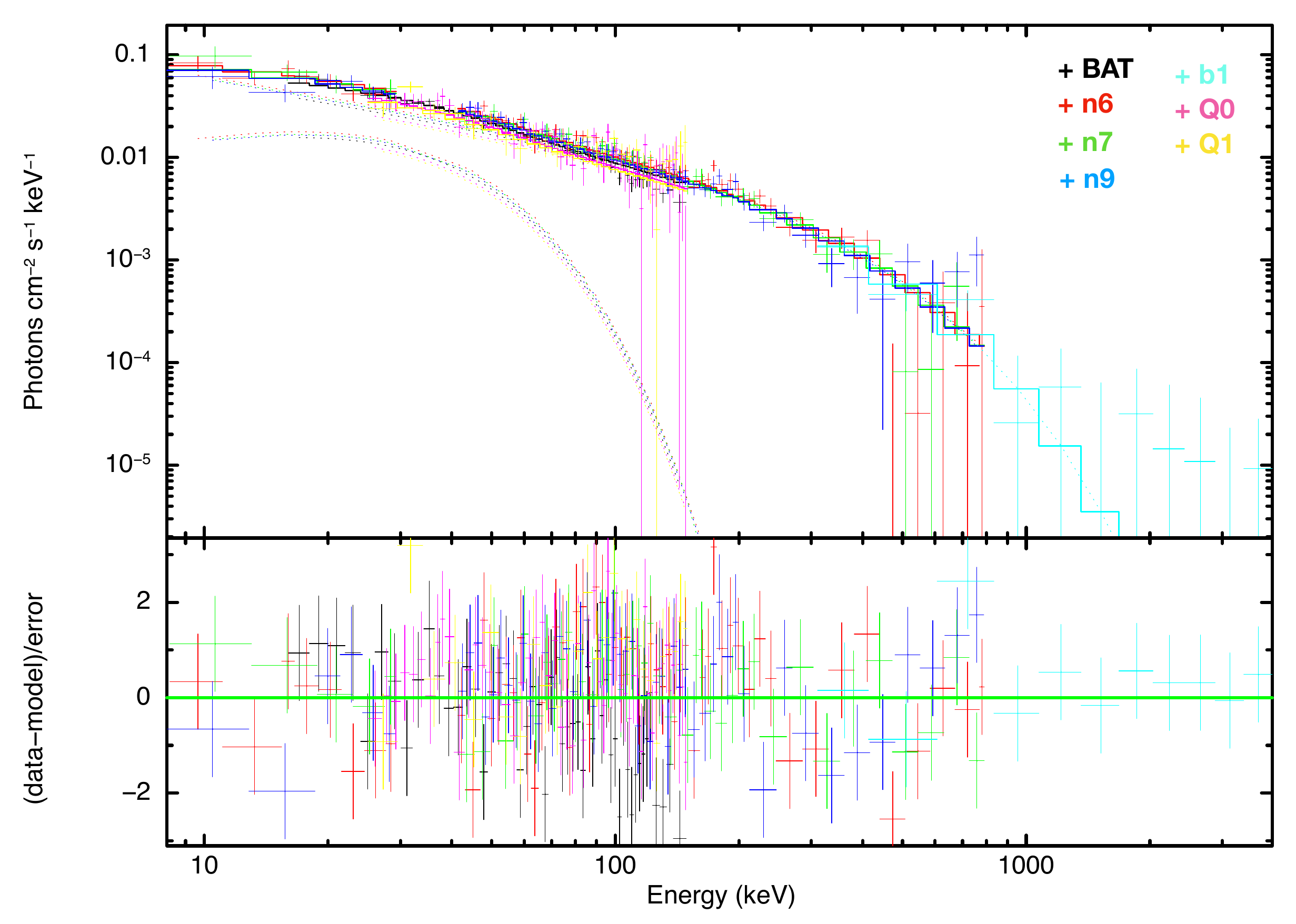}
\end{minipage}%
\begin{minipage}{.5\textwidth}
  \centering
  \includegraphics[width=1.0\linewidth]{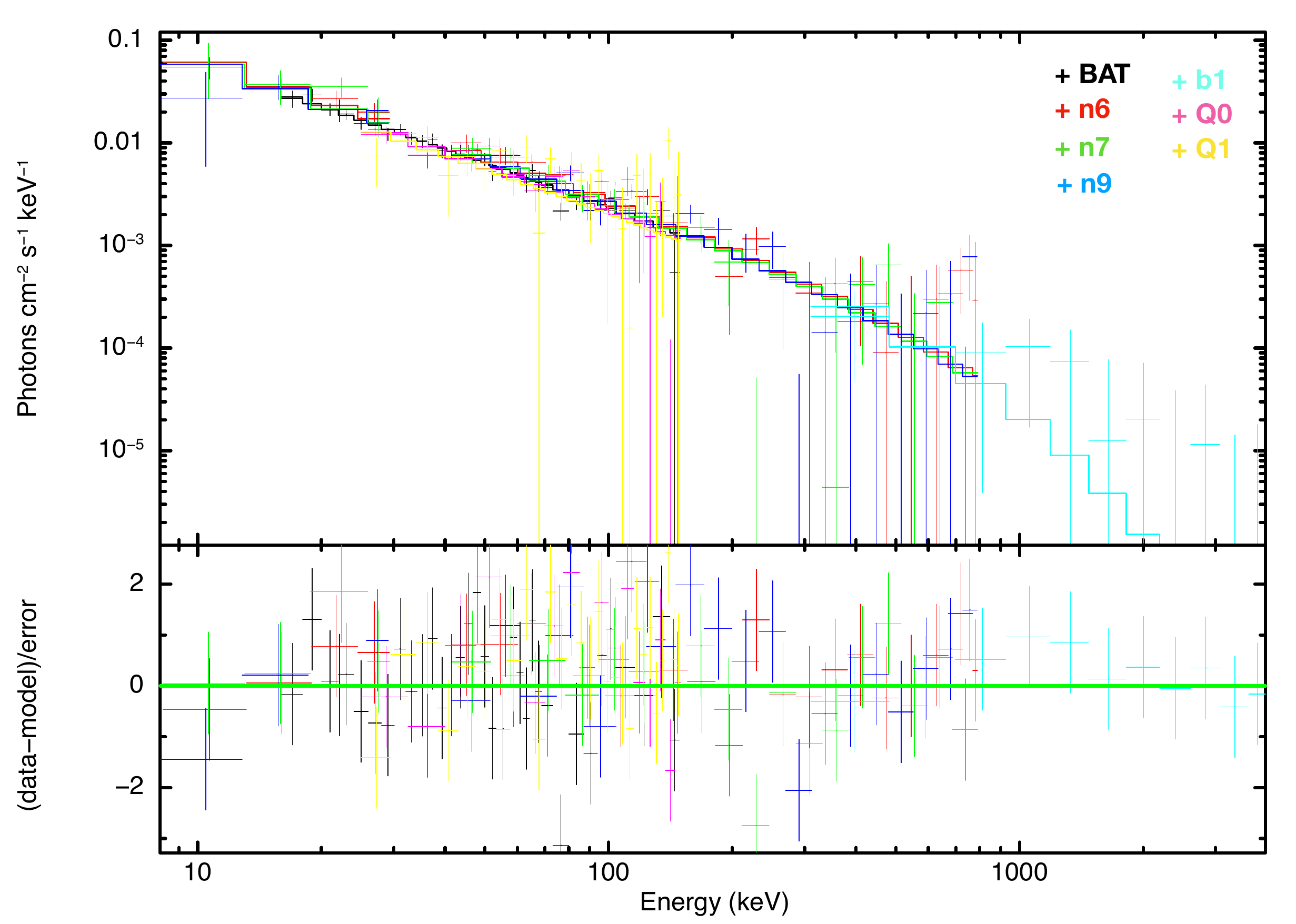}
\end{minipage}
\caption{Counts spectra for the time integrated analysis of the first (left) and second (right) emission episodes including data obtained from {\it AstroSat}/CZTI, {\it Fermi}/GBM and {\it Swift}/BAT are shown best fit model in the upper panels. The residuals obtained for the respective spectral fits are shown in the lower panels.}
\label{fig:160821A_spectral_fit_czti}
\end{figure*}

We have performed joint spectrum fitting of GRB 160325A using {\it AstroSat}/CZTI, {\it Fermi}/GBM and {\it Swift}/BAT for each episode separately. In the calibration study using the Crab Nebula spectrum observed during 31st March 2016, it was found that quadrants Q2 and Q3 showed significant deviations in the estimate of constant of normalisation. Therefore, we have used only quadrants Q0 and Q1 in the current spectral analysis. 
For this task, the CZTI spectrum and response matrix files of each quadrant are extracted using standard CZT modules\footnote{http://astrosat-ssc.iucaa.in/uploads/czti/CZTI$\_$level2$\_$software$\_$userguide$\_$V2.1.pdf}: GRB and background interval spectrum files are obtained using cztbindata and response is generated using cztrspgen. To account for the inter-instrument calibration, a constant of normalisation is multiplied to the model for each CZTI quadrant while keeping that for the BAT detector to be unity and those for the {\it Fermi} detectors at the respective effective area correction values found in section 3.1. We find the relative normalization of Q0 is $0.91^{+0.03}_{-0.03}$ and $0.87^{+0.04}_{-0.04}$ for Q1. The spectral models blackbody+cutoffpl and cutoffpl are fitted with the data of all three instruments, for the two episodes respectively. The counts spectra along with the fit residuals %
of both the episodes are shown in Figure~\ref{fig:160821A_spectral_fit_czti}. 
Values of the best-fit parameters are listed in Table~\ref{tab:joint_spectrum}. 

\begin{table*}
\begin{small}
\caption{Results of time integrated joint spectral analysis using BAT, {\it Fermi} and {\it AstroSat} data. }
\label{tab:joint_spectrum}
\begin{center}
\begin{threeparttable}
\begin{tabular}{p{1.0cm}ccccccc}
\hline 
Episode & Time interval   & $kT$      & - $\alpha$	& $E_{c}$   & $F_{T}/F_{NT}^{*}$   &  Reduced-$\chi^{2}$
\\
$\#$ & (s)     & keV       &           	& keV         &   &   
\\ \hline \hline
1 & -3 - 21	& $10.64_{-0.62}^{+0.70}$  & $0.34_{-0.07}^{+0.07}$	& $225_{-26}^{+27}$	& $\sim10 \%$ & 1.06
\\
2 & 30 - 52.5	& --  & $1.24_{-0.07}^{+0.07}$	& $539_{-174}^{+333}$  &  --    & 1.01
\\ \hline \hline
\end{tabular}
\end{threeparttable}
\smallskip\scriptsize
\begin{tablenotes}[flushleft]
  \item[*] $F_{T}/F_{NT}$ represents the percentage of ratio of thermal flux to the non-thermal flux in $10-1000$ keV range. Whereas, in the polarisation measurement energy range $100-400$ keV this ratio is found to be $\sim 18 \%$.
\end{tablenotes}
\end{center}
\end{small}
\end{table*}

\subsection{Polarisation Measurements}
CZTI is experimentally verified for polarisation measurement capability in 100-400 keV energy range for on-axis sources \citep{vadawale2013prospects}. At energies of a few hundreds of keV, the dominant mode of photon interaction is Compton scattering. Linear polarisation signature in the GRB is estimated through the non-uniform azimuthal distribution of the Compton-scattered photons. First, on-board verification of CZTI X-ray polarimetry measurement capability was done by the measurement of the polarisation of Crab nebula and pulsar in 100-380 keV energy range \citep{vadawale2018phase}. GRB 160325A has been found to be polarised at $> 2 \sigma$ confidence level with a high polarisation fraction of $59\pm28\, \%$ for the whole burst \citep{2019ApJ_Chattopadhyay}. In this work, we have adopted the same analysis procedure as described in the above references, but have conducted the polarisation analysis of the two emission episodes separately.

\begin{figure}
	\includegraphics[width=\columnwidth]{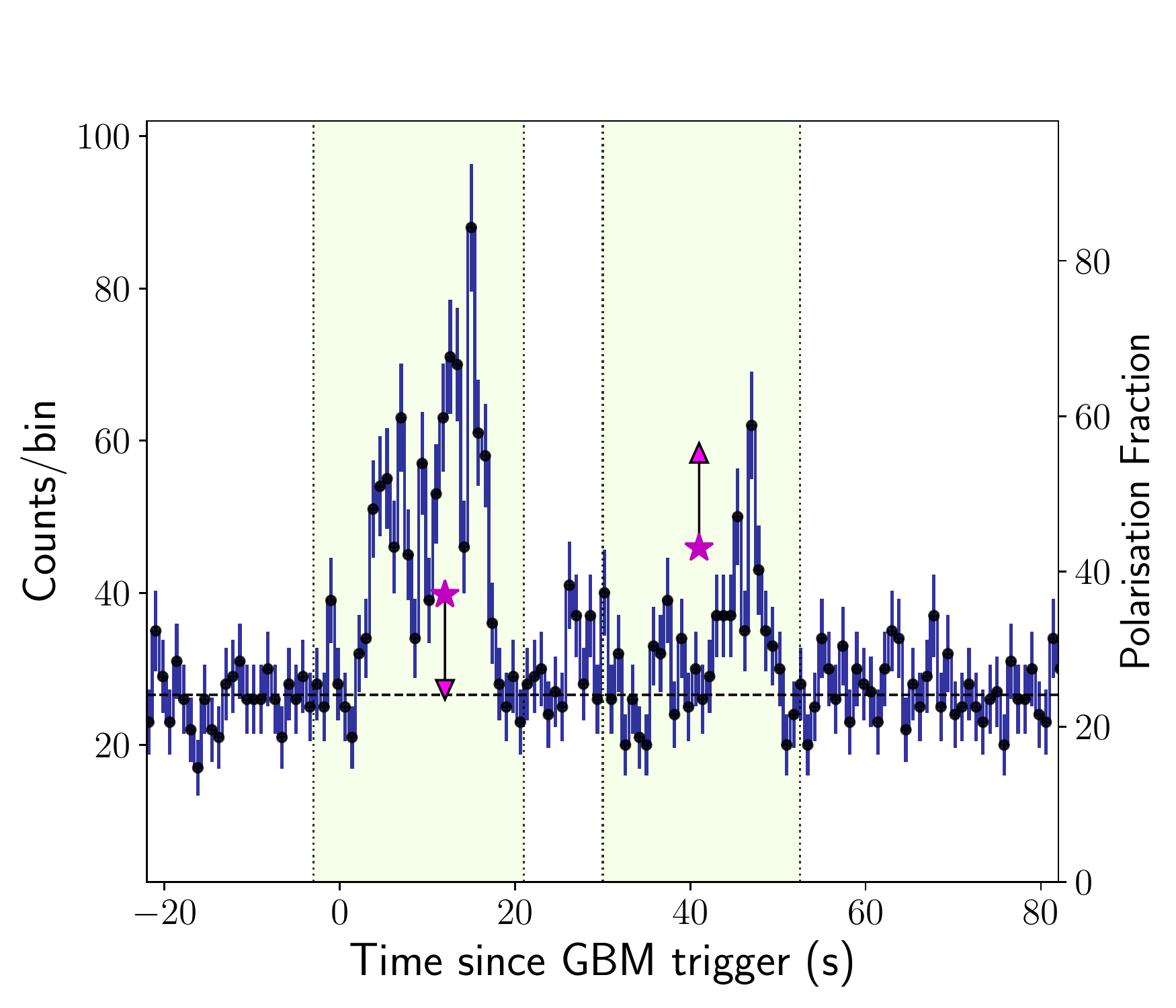}
    \caption{Compton lightcurve of GRB 160325A for 0.8 s binsize is shown. The time intervals of polarisation measurements are shown in two shaded regions for two distinct episodes of GRB. The mean background is shown with a black horizontal dashed line. The upper and lower limits of the polarisation fraction measured at $1.5\, \sigma$ confidence level in the two emission episodes are marked in magenta star.}
    \label{fig:Compton_lightcurve}
\end{figure}

The Compton events are identified by first  selecting the double-pixel events that occurred within a 40$\mu$s time window, and secondly, by filtering those double-pixel events against the various Compton kinematics criteria \citep{chattopadhyay2014prospects}. 
The Compton lightcurve of GRB 160325A in 100-380 keV energy range is shown in Figure~\ref{fig:Compton_lightcurve}. The burst is clearly seen in terms of Compton events. The mean background rate of Compton events is 21 counts/s. The azimuthal scattering angle distribution of events is computed for the GRB prompt emission interval as well as for background intervals before and after the prompt emission. The background corrected GRB azimuthal distribution is obtained after subtracting the combined pre- and post-burst background azimuthal distribution. 

The azimuthal scattering angle distribution of the Compton scattered photons are expected to be of the sinusoidal form:
\begin{equation}
    C = A \; cos[2(\phi-\phi_{0}+\pi/2)]+B
	\label{eq:sin_azi}
\end{equation}
where, $\phi_{0}$ represents the polarisation angle in detector plane, parameters $A$ and $B$ are related to the modulation factor ($\mu$) such that $\mu = A/B$. The background subtracted and geometry-corrected azimuthal distribution is then fitted by this sinusoidal function. As the polarisation angle obtained from fitting the modulation curve is in CZTI detector plane, it is then converted into the celestial frame of reference and this is reported in the paper. The polarisation fraction is estimated as the ratio of the observed modulation factor to the modulation factor expected for $100\%$ polarised radiation, $\mu_{100}$. The $\mu_{100}$ simulations are performed in Geant4 with a full mass model of {\it AstroSat} and CZTI for many polarisation angles. The simulation is performed with $10^{8}$ photons generated with the same spectral shape and incoming direction as that of the GRB.

For GRB 160325A, the polarisation measurement is performed for the selected time intervals (shaded region in Figure~\ref{fig:Compton_lightcurve}) corresponding to the two episodes of the burst. The number of Compton events during the two emission episodes are $\sim 690$ and $\sim 370$, respectively. During the first interval, the azimuthal angle distribution shows a very low modulation, making it consistent with unpolarised emission. The polarisation angle is also poorly constrained during that interval with a large uncertainty, signifying either burst is inherently unpolarised (Appendix B) or the polarisation fraction is below the CZTI sensitivity. Therefore, here we report an upper limit on polarisation fraction as $37\%$ and $50\%$ at $1.5\sigma$ ($86\%$) and $2\sigma$ ($90\%$) confidence level respectively, for one parameter of interest. The low polarisation fraction across the emission episode can be due to (i) varying polarisation angle as a function of time within the episode, despite high polarisation being high \citep{gotz2009variable,zhang2019detailed,burgess_etal_2019,2019ApJ_sharma} or (ii) the observed emission within $1/\Gamma$ is inherently weakly polarised (\citealt{Toma_etal_2009}, also see section \ref{first_pulse}). 

\begin{figure*}
\centering
\begin{minipage}{.499\textwidth}
  \centering
  \includegraphics[width=1.0\linewidth]{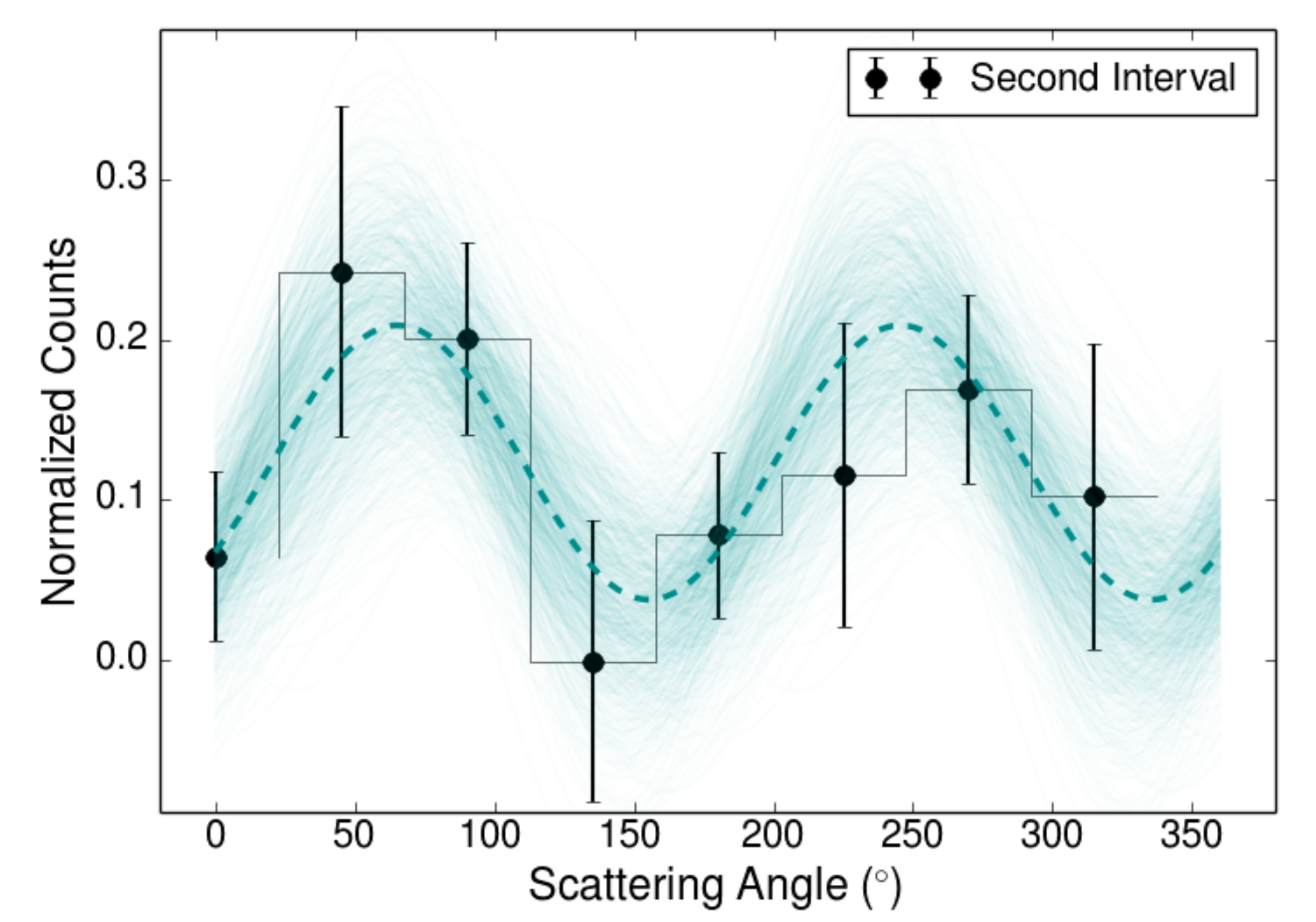}
\end{minipage}%
\begin{minipage}{.369\textwidth}
  \centering
  \includegraphics[width=1.0\linewidth]{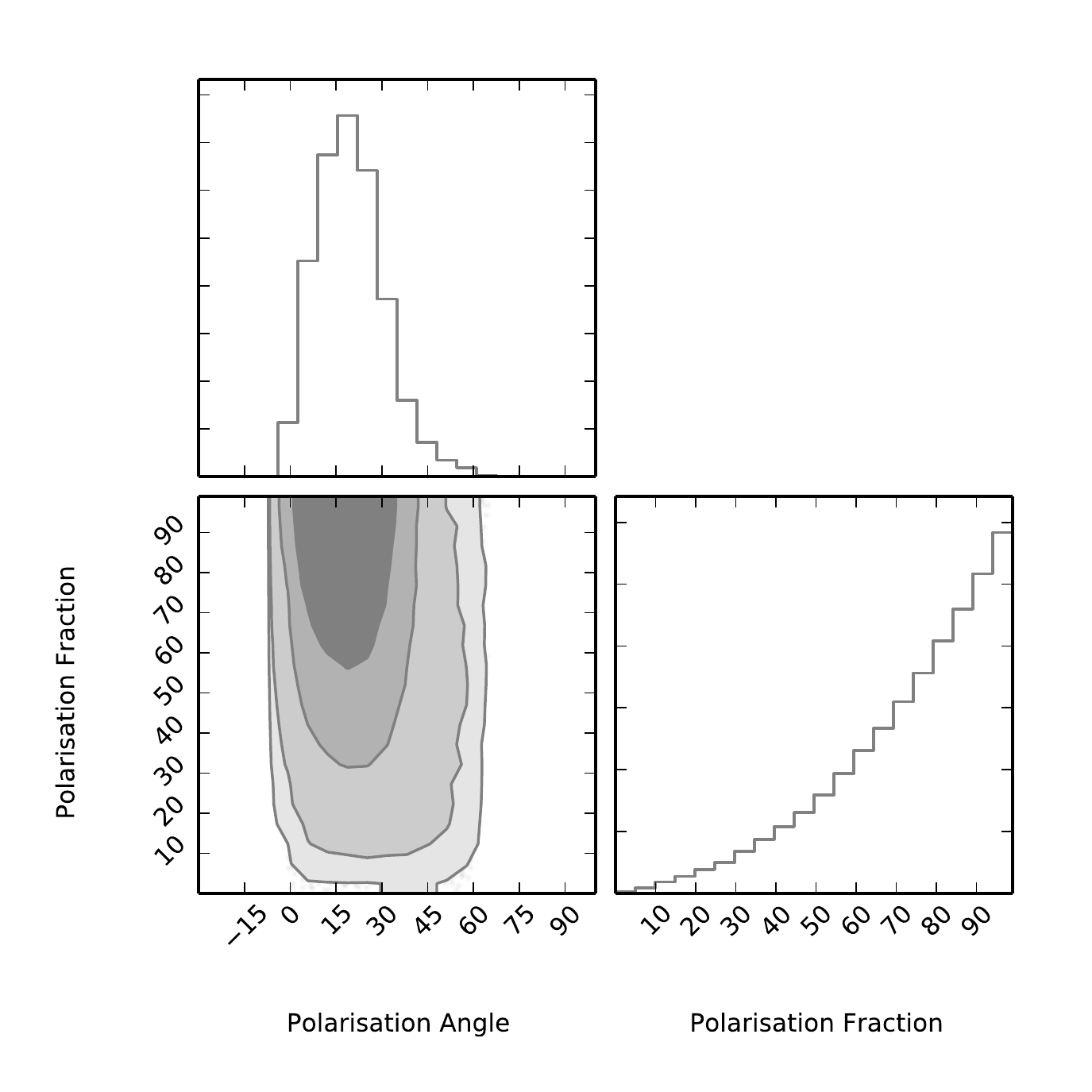}
\end{minipage}
\caption{Azimuthal scattering angle distribution (left) and contour plot (right) of polarisation angle and fraction with $68\%$, $90\%$, $99\%$ and $99.9\%$ (top to bottom) confidence levels for two parameters of interest obtained 
for the second episode are shown.}
\label{fig:Interval2_pol}
\end{figure*}

Despite the small number of Compton events, the second emission episode, on the other hand, is found to be highly polarised at $>3\sigma$ confidence level (Figure~\ref{fig:Interval2_pol}) for two parameters of interest. This is because for on-axis GRB detection, the polarisation sensitivity of the instrument is high \citep{chattopadhyay2014prospects}. In Figure~\ref{fig:Interval2_pol}, 2-D contours of confidence intervals of $68\%$, $90\%$, $99\%$ ($3 \, \sigma$) and $99.9\%$ ($3.7\, \sigma$)   of polarisation fraction and polarisation angle are shown\footnote{The method of Monte Carlo simulations to obtain the 2-D histogram of PF and PA is described in \cite{2019ApJ_sharma}}. 
We find the lower limit of polarisation fraction to be $>43\%$ (Appendix D), and the polarisation angle to be $18\pm15^{\circ}$, at $68\%$ confidence level for two parameters of interest. The results of the polarisation measurements are listed in Table~\ref{tab:polarisation}. Similar varying polarisation fraction across different emission episodes has been previously reported in case of GRB 041219A \citep{gotz2009variable}.

\begin{table*}
\begin{small}
\caption{Results of polarisation measurements of GRB 160325A}
\label{tab:polarisation}
\begin{center}
\begin{threeparttable}
\begin{tabular}{p{2.0cm}ccccccc}
\hline 
 Time Interval	& Energy Band & Compton events		& PF ($\%$)		& PA  ($^\circ$)      & PF ($\%$)		& PA  ($^\circ$)
\\
   (s)        	& (keV)         &     	& (1.5$\sigma$ C.I.)	    &                    & (2$\sigma$ C.I.)	    & 
\\ \hline \hline
 -3 - 21.0 	   & 100-380	  &$\sim 690$     & $<37$         & -- & $<50$ & --
\\ 
 30.0 - 52.5 	& 100-380     &$\sim 370$     & $>43$        & $18 \pm 15$ & $>20$ & $18_{-20}^{+21}$
\\ \hline \hline
\end{tabular}
\smallskip\scriptsize
\begin{tablenotes}
  \item[]\textit{Note}: C.I. notation is used for Confidence interval. The upper limit of PF is reported for one parameter of interest during first episode. For second episode, the lower limit on PF is reported with PA values for two parameters of interest from the 2-D contour plot.
\end{tablenotes}
\end{threeparttable}
\end{center}
\end{small}
\end{table*}

\section{Afterglow}
\label{ag}
\begin{figure}
	\includegraphics[width=\columnwidth]{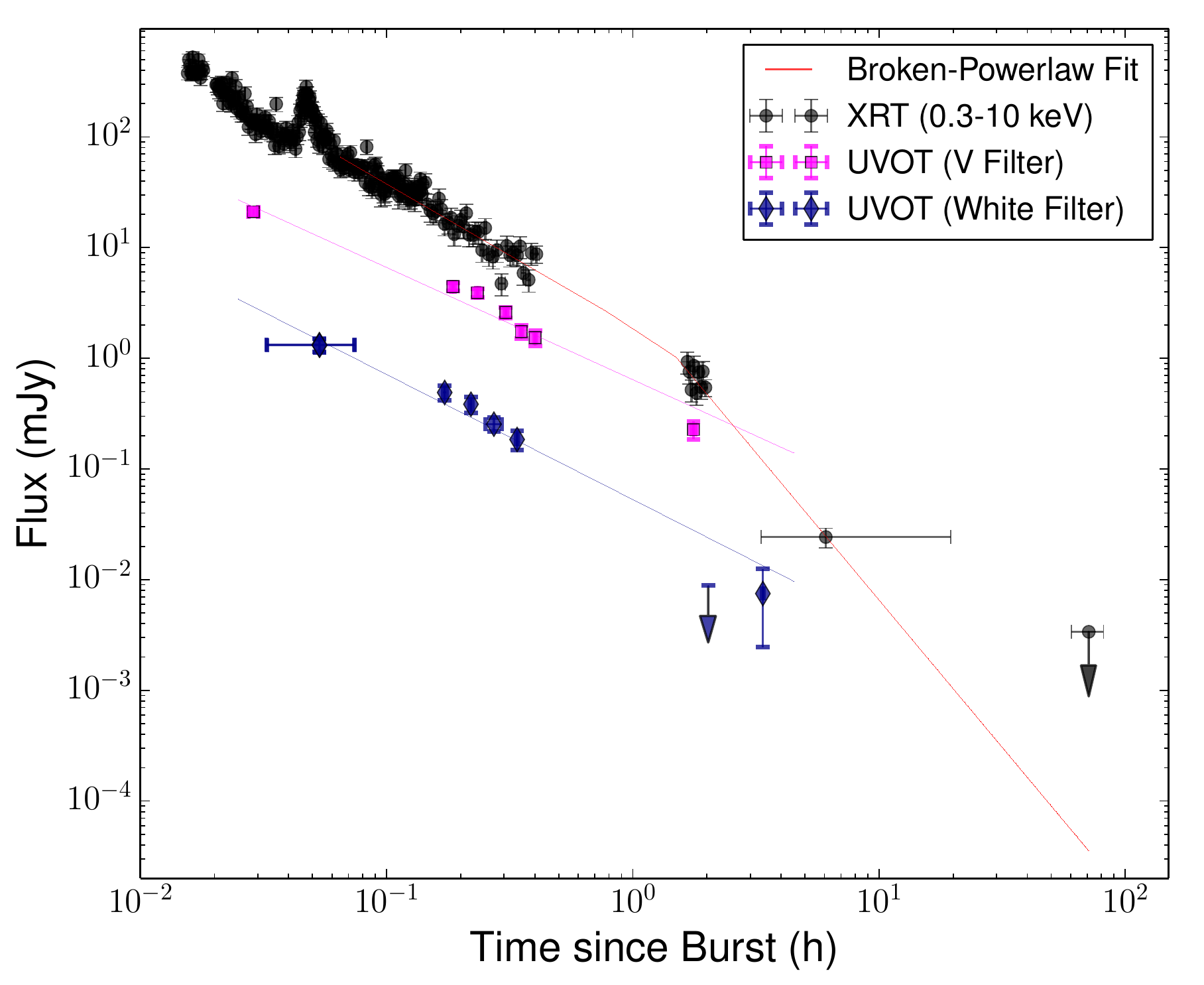}
    \caption{{\it Swift} afterglow flux evolution is shown indicating the presence of a break in the flux lightcurve. {\it Swift}/XRT flux lightcurve is shown in black circle and {\it Swift}/UVOT: v and white filters flux lightcurve is shown in magenta square and blue rhombus, respectively. The broken power-law fit to {\it Swift}/XRT lightcurve is shown in a solid red line.}
    
    \label{fig:figure_xrt_lc}
\end{figure}

GRB 160325A was observed by {\it Swift}/XRT $\sim 60$ seconds after the burst trigger time in 0.3-10 keV energy range. {\it Swift}/XRT online repository\footnote{http://www.swift.ac.uk/xrt$\_$curves/00680436/} is used for generating the flux light curve and spectral files \citep{evans2009methods}. In Figure~\ref{fig:figure_xrt_lc}, the energy flux light curve is shown. A flare as well as a jet break are evident in the XRT data when fitted with a single power law. The optical afterglow is also detected by the {\it Swift}/UVOT detector. Photometric analysis is performed by taking a circular aperture of $3{\arcsec}$ around the optical afterglow position. The fluxes obtained are corrected for Galactic extinction in the direction of the burst using the model described in \cite{cardelli1989relationship}. The light curves in v and white filters flux lightcurve hint towards the presence of a jet break (Figure~\ref{fig:figure_xrt_lc}), as the last few points do not follow the power-law fit to the earlier segment.

We find that the flux light curve after the flare is best fitted by a broken power law function as shown in Figure \ref{fig:figure_xrt_lc}. The pre-break and post-break power law indices $\phi$ and $\phi_{1}$ are $1.30 \pm 0.04$ and $2.66 \pm 0.19$ respectively with the break occurring at a time $t_{b} = 1.38 \pm 0.15$ h.
The X-ray spectrum in the time interval (260-1460 s) namely the region after the flare and before the break is best fitted by a power law with photon index $\zeta = 2.28 \pm 0.33$. The corresponding energy spectral index is $\beta \equiv 1-\zeta$. 

The  fireball  model  predicts that the afterglow emission is produced when the outflow crashes into the circumburst medium resulting in external shocks. Electrons accelerated in the shocks then emit via synchrotron emission. By assessing the closure relation between the temporal decay index $\phi$ and spectral index $\beta$ \citep{racusin2009_jet}\footnote{Convention is $F_{\nu} \; \propto \; t^{-\phi} \; \nu^{-\beta}$}, we identify that the observed synchrotron emission belongs to the spectral segment, $\nu > \, \rm max(\nu_m,\nu_c)$, of a fast cooling synchrotron emission produced in an external shock propagating in a uniform circumburst medium \citep{sari1998}. The index of the power law energy distribution of the accelerated electrons is $p = 2 \times \beta = 2.56 \pm 0.65$.  

We then estimated the kinetic energy of the jet to be $E_{k,iso} = 1.34 \times 10^{54}\, \rm erg$ by using the equation (17) in Appendix A of \citealt{wang2015bad} (or equation (11) in \citealt{zhang2007grb}) in which the following assumptions were made: the fraction of post shock thermal energy in magnetic fields, $\epsilon_B = 0.1$ and in electrons, $\epsilon_e =0.1$; negligible inverse Compton scattering 
and a constant ambient number density, $n=1 \, cm^{-3}$. The radiation efficiency of the prompt emission of the GRB is given by $\kappa = E_{\gamma,iso}/(E_{\gamma,iso} + E_{k,iso})$ and is estimated to be $11\, \%$. 

The jet break time from observational data is used to calculate the opening angle of the conical blast wave by the following formulation:
\begin{equation}
    \theta_{j} = 0.070 \; rad \; \left(\frac{t_{b}}{day}\right)^\frac{3}{8} \; \left(\frac{1+z}{2}\right)^{-\frac{3}{8}} \; \left(\frac{E_{k,iso}}{10^{53} \; erg}\right)^{-\frac{1}{8}} \; \left(\frac{n}{0.1  \; cm^{-3}}\right)^\frac{1}{8}
	\label{eq:jet_break}
\end{equation}

which is described in \cite{sari1999_jets,wang2015_break}. For redshift $z = 2$, the jet opening angle is found to be $\theta_{j} = 1.2^{\circ}$.

\section{Discussion}
\subsection{Observing geometry}
\label{obs_geometry}
The orientation of the jetted outflow with respect to the observer is critical in inferring the radiation mechanism from polarisation measurements. Since spatially resolved polarisation measurements are not made, any polarisation arises due to some asymmetry in the distribution of the polarisation vectors within the viewing cone of the jet.
For example in the case of synchrotron emission produced in random magnetic field and of inverse Compton scattering, the polarisation is expected to be very low ($<5 \%$) when viewed from within the jet cone ($< \theta_j$). Thus, in the case of top-hat jets, a net polarisation is obtained only in certain viewing geometries where the jet is observed off-axis ($> \theta_j$) and when the symmetry of emission across the line of sight is broken. On the other hand, synchrotron emission produced in ordered magnetic field is expected to be highly polarised even when viewed on-axis within the jet cone. 
Thus, in addition to the spectral information, polarisation measurement can be used to decipher the radiation process once the viewing geometry is known. 

\cite{kumar2000afterglow}, \cite{granot2002off}, \cite{Van_Eerten_etal_2010} have shown that the temporal evolution of the afterglow flux significantly varies when viewed within the jet cone or when viewed off-axis. In case of a top-hat jet, when viewed off-axis, a fast rise in energy flux is observed with time and the peak of energy flux is observed nearly when the viewing cone of $1/\Gamma$ becomes equal to the off-axis observer angle. On the other hand, when viewed on-axis, we observe a decreasing energy flux and a break signifying a steeper decrease in flux evolution is observed when the viewing cone of $1/\Gamma$ becomes equal to the jet opening angle. After $1/\Gamma > \theta_j$, the jet may start to expand laterally, then in that case, the flux is expected to decrease such that $F \propto t^{-p}$ where $p$ is the electron power law index \citep{Rhoads_1999,Van_Eerten_etal_2010,Van_Eerten_etal_2012,Ryan_etal_2015}.   

In case of GRB 160325A, the afterglow emission is observed as early as 60 s post GRB trigger.  (Figure~\ref{fig:figure_xrt_lc}). 
After the break, $t_{b} = 0.06 \pm 0.01$ day, the flux decreases as a power law whose index is found to be consistent with the electron power law index (section \ref{ag}). 
Thus, this break observed in the flux evolution is associated to the jet break \citep{racusin2009_jet}. This temporal behaviour of the afterglow emission suggests that the GRB jet is pointed towards the observer such that the line of sight lies within the jet cone. With this understanding of the viewing geometry, we now decipher the radiation process giving rise to the observed two emission episodes  in the subsequent subsections. 

\subsection{Jet composition: Baryon dominated outflow with mild magnetisation}
\label{jetcomp}
Previously, in cases like GRB 080916C \citep{zhang2009evidence} and GRB 160625B \citep{zhang2018transition}, the presence or the absence of the thermal component in the spectrum of the GRB 
has been used as an indicator to infer if the jet is dominantly composed of Poynting flux or not. In the case of GRB 160625B, the initial pulse was found to possess a strong thermal component and the second emission pulse that was observed after a long quiescent period of $\sim 180$ s was found to possess a non-thermal spectrum, best modelled using a Band function alone. In \cite{zhang2018transition} and \cite{Li2019}, these spectral observations were used to infer that the burst has undergone a transition from a pure fireball to Poynting flux dominated outflow. Here in the case of GRB 160325A, we also find a similar spectral transition from a thermal component and a non-thermal component inconsistent with synchrotron emission, detected in the first episode to that of only a non-thermal component consistent with synchrotron emission in the second episode. The variation of the polarisation fraction from that of a relatively low ($<37\%$) in the first episode to that of a high ($> 43\%$) in the second episode, also tend to support the hypothesis of the jet transiting from a fireball to a Poynting flux dominated outflow or not. 
In the below discussion, we investigate whether the second emission episode originates in a Poynting flux dominated outflow.

The radiation efficiency of the burst is estimated to be only $11\%$ which is very low (section \ref{ag}). In Poynting flux dominated outflow models, radiation efficiencies can range between $3 -50\%$ \citep{Giannios2006,Drenkhahn_Spruit2002,Zhang_Yan_2011}. However, in these models, we note that low radiation efficiencies $\sim 3-20\%$ are expected only when the dominant part of the observed emission comes from the photosphere. 
For example, in \cite{Giannios2006} the dominant observed emission is expected to be coming from the photosphere with most of the Poynting flux dissipating below it and thereby leading to a non-thermal like emission spectrum. In this case, the photospheric luminosity is found to be at the maximum $\sim 20\%$ of the total burst luminosity.  
Figure 4b in \cite{Drenkhahn_Spruit2002} shows that the radiation efficiency of the non-thermal emission (the emission produced in the optically thin region above the photosphere) is expected to be $<20\%$ for moderately magnetized outflows such that $\sigma_0$\footnote{The ratio of Poynting flux luminosity to fireball luminosity} is only a few tens. In such cases, a strong thermal component (i.e $F_T \, \ge \, F_{NT}$) is expected to be observed in the spectrum. This, however, is not observed in both the emission episodes of GRB 160325A. 
Also, a dominant photospheric emission is expected to give rise to low polarisation fraction which is not observed in the second emission pulse. 
If we associate the non-thermal spectrum of second emission pulse to the radiation from a dissipative photosphere model, irrespective of the dissipation mechanism, \cite{Lundman_etal_2018} have shown that polarised emission can be expected from the photosphere if the jet is significantly magnetized, thereby producing synchrotron photons in regions close to the (both above and below) photosphere. Under most favourable conditions a maximum limit of PF $\sim  50\%$ can be achieved, however, significant polarisation is expected only in the energy range much lower than $100\, \rm keV$. This is in contradiction to that observed in the second emission episode where high polarisation is observed in $100-380 \, \rm keV$ range. This thereby suggests that the highly polarised non-thermal emission observed in the second emission episode originates in the optically thin region of the outflow. 

In the ICMART model \citep{Zhang_Yan_2011}, a highly magnetized outflow ($\sigma_0 \gg 1$) with a highly efficient radiation close to $\sim 50\%$ is expected. At the same time, within each ICMART emission event, the polarisation is expected to decrease as the magnetic field configuration gets more randomized at the end of the Poynting flux dissipation. As a result, on average polarisation fraction when measured across multiple emission episodes is expected to be only a few percent. 
Thus, we find that the possibility of the second emission episode to be produced within a Poynting flux dominated outflow is less likely.

On the other hand, low radiation efficiency is expected in the internal shock model wherein due to variations in the outflow, a fast moving shell crashes into a slow moving one, thereby dissipating only the differential kinetic energy of the shells \citep{Mochkovitch_etal_1995,Kobayashi_etal_1997,Daigne_Mochkovitch1998}. In internal shocks, random magnetic fields are produced, which is largely expected to result in very low polarisation fraction when viewed on-axis \citep{Granot_2003,Toma_etal_2009}.  
But it was suggested by \cite{Waxman_2003} that during the prompt emission, when a narrow top hat jet ($\theta_j \sim 1/\Gamma$) is viewed along the edge such that the line of sight lies along $\theta_j + 1/\Gamma$, a high polarisation fraction can be expected even in the case of random magnetic fields. 
Later, in the detailed study done in \cite{Granot_2003}, it was shown that only if there exists an anisotropic field configuration behind the shock front such that the magnetic field lines ($B_{||}$) are parallel to the direction normal to the shock, then the average PF across multiple pulses within the burst can be $\sim 35\% - 62\%$. 
However, the polarisation measurements done during the afterglow observations show low values of PF ($<10\%$) \citep{Wijers_1999_990510_afterglowpol,Covino_2002_011211,Greiner_2003_030329_opticalpol,King2014_131030A_opticalpol,Gorbovskoy_2016_opticalpol,troja2017significant,Laskar_etal_2019}. 
This in turn suggests that the magnetic fields generated in the relativistic shocks are isotropically distributed. Thus, the $B_{||}$ field configuration appears unlikely to occur in internal shocks. 
With this understanding, we find that the possibility of second emission episode to be the radiation from internal shocks produced in optically thin region of a purely baryonic outflow is also less likely, as it cannot account for the high polarisation fraction observed in the second emission episode.

From the above discussions,
we, thus, find it more reasonable to envisage a scenario such that the outflow is baryon dominated with a subdominant Poynting flux component ($\sigma_0 \ll 1$) which remains passive through out the burst duration (see section \ref{sp_inter}) and dissipation of the kinetic energy of the jet happens via internal shocks. The subdominant Poynting flux component can result in some net ordered magnetic fields at the dissipation site. Further discussions and inferences regarding the GRB are presented within this scenario. We also note that in the calculations used for inferring the outflow dynamics, the contribution of Poynting flux component is negligible. 

\subsection{Intermittent emissions} 
\label{int_em}
There exists a low level emission for a duration of $\sim 9$ s between the two emission episodes.
If we assume the central engine is continuously active throughout the burst and, in the scenario of internal shocks, consider 
the time interval between the start of the first and the second episodes, $\Delta t_{obs} \sim 40\, \rm s$, to represent the variability timescale, 
$t_v$, of flow that generates shocks responsible for the second emission episode, then the associated length scale works out to be $c \, t_v  \sim 10^{12} \, \rm cm$ where $c$ is the speed of light. In general, the variations in the outflow are expected to happen on several possible length scales for example,  the size of the central engine: a stellar mass black hole or a magnetar ($10^{6-7} \, \rm cm$ ), the nozzle radius of the jet ($10^{6-9}\, \rm cm$, \citealt{iyyani2013variable,Gao_Zhang2015,Iyyani_etal_2016}), the radius of the accretion disc (three times the Schwarzschild radius) or the radius of recollimation shocks within the stellar core ($10^{9-10}\, \rm cm$, \citealt{LopezCamara_Morsony2013,Mizuta_Ioka2013}). The length inferred above is quite large and variability in the outflow in such scales is generally not expected.
The second emission episode is therefore better explained as a fresh injection of energy from the central engine after nearly $9 \, \rm s$ 
of mild activity or quiescence. 
If the central engine is in fact dormant during the interval between the emission episodes, then the soft emission observed during this period may be related to the emission coming from the high latitudes (curvature effect, \cite{Dermer2004,Uhm_zhang2015}).

\subsection{First emission episode}
\label{first_pulse}
The spectrum of the first emission episode reveals a blackbody component and a non-thermal component (cutoff power law) which possesses a low energy power law index that lies largely above the line of death of synchrotron emission. The thermal emission is found to contain only $\sim 9\%$ of the total observed flux in the energy range $10-1000\, \rm keV$. The hard spectral slopes are incompatible with optically thin synchrotron emission 
from above the photosphere.
Optically thin inverse Compton scattering
is also unlikely as the source of this emission, since very high energy photons ($\ge \, \rm MeV$) are expected \citep{ghirlanda2003extremely}, while here a spectral cutoff in the spectrum at a few hundred keV is observed and no significant emission is seen above $1.1 \, \rm MeV$. Thus, we interpret the observed non-thermal emission to be originating from the photosphere itself.
Almost no polarisation is expected from the photosphere when viewed on-axis \citep{Lundman_etal_2014,Lundman_etal_2018}. The polarisation measurement of the emission episode places an upper limit of $<37\%$ at $1.5\, \sigma$ confidence level of one parameter of interest. This is consistent with our interpretation of subphotospheric dissipation model involving quasi-thermal Comptonization (see also \citealt{Liang_etal_1997,Ghisellini_Celotti1999}). 

The non-smooth and broad shape of the spectrum originating from the photosphere indicates localised dissipation happening below the photosphere \citep{Pe'er_Waxman2005,pe2006observable,iyyani2015extremely} when the outflow is in the coasting phase. Since the dissipation happens below the photosphere at some moderate/ large optical depth ($\tau \gg 1$), the electrons attain a steady state and remain sub-relativistic due to a balance between heating and cooling \citep{pe2005peak}. In such a scenario, the thermal photons produced near the central engine, when advected into the dissipation site, get upscattered by the subrelativistic electrons. Due to a large number of subsequent inverse and direct Compton scatterings before the Comptonized photon field gets decoupled at the photosphere, a cutoff peak is formed at higher energies via Comptonisation accompanied by the adiabatic cooling due to outflow expansion. 

\begin{figure}
	\includegraphics[width=\columnwidth]{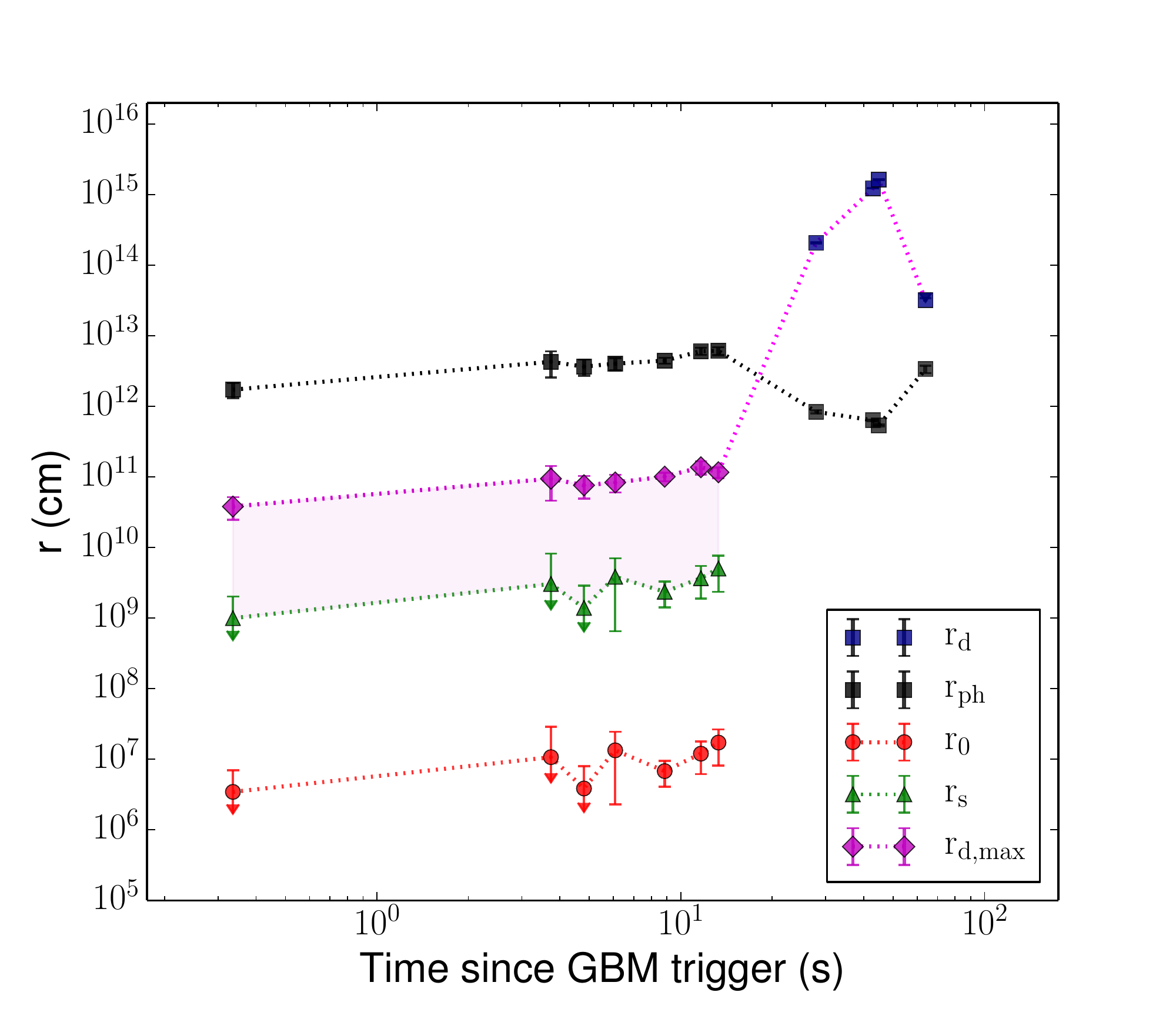}
    \caption{The temporal evolution of the outflow parameters of the jet during the first and second emission episodes are shown. The nozzle radius, $r_{0}$ (red circles), saturation radius, $r_{s}$ (green triangles) and photosphere radius, $r_{ph}$ (black square) are shown. 
    The maximum possible dissipation radius, $r_{d, max}$ (magenta diamond) in case of subphotospheric dissipation in the first emission episode and the dissipation radius, $r_d$ (dark blue squares) estimated in the second emission episode considering optically thin internal shocks are also shown.  
    }
    \label{fig:radius}
\end{figure}

Here we compute the outflow jet properties and estimate an upper limit to the  dissipation radius by following the methodology adopted in \cite{pe2007new} (also see \citealt{Gao_Zhang2015}) and \cite{iyyani2015extremely} respectively. For this, we associate the observed blackbody emission to the seed thermal photons of energy ${\cal{E}}_i = 2.7\, kT$ that are advected from the central engine, and the cutoff peak energy to the Comptonised peak, ${\cal{E}}_c$. 
We find the outflow parameters of the jet (Figures \ref{fig:radius} and \ref{fig:gamma}) on average to be the following: Lorentz factor of the jet at the photosphere, $\Gamma_{ph} =\, 310$, the photospheric radius, $r_{ph} \sim 4.3 \times 10^{12}\, \rm cm$, saturation radius, $r_s \sim 2.9 \times 10^{9}\, \rm cm$ and the nozzle radius of the jet, $r_0 \sim 9.6 \times 10^{6} \, \rm cm$. The possible range of the dissipation radius ($r_d$) is between $r_s$ and  $r_{d, max}$ which is found to be $\sim  9.3 \times 10^{10}\, \rm cm$ using the equation 2 and 3 in \cite{iyyani2015extremely}.  The factor, $f$ which considers if the electrons are in the Thompson regime or not, is assumed to be $\sim 2$ in the calculations.  

It can be envisaged that due to oblique or collimation shocks that occur as the jet pierces through the stellar core, the dissipation of the kinetic energy of the jet can happen at such a small radius below the photosphere \citep{LopezCamara_Morsony2013,Duffell_Macfadyen2015,Mizuta_Ioka2013}. The variability of the Lorentz factor of the jetted outflow in this case can be expected to occur at a time scale of $t_v = r_0 \theta_j/c$ 
\citep{Rees_Meszaros2005}. This variability can cause the internal shocks to occur at a radius, $r_d = 2\Gamma^2 c\, t_v = 4 \times 10^{10}\, \rm cm$. This estimate is consistent with the upper limit of the dissipation radius that was made using the Comptonized peak of the observed spectrum. This corresponds to an optical depth of $\tau = r_{ph}/r_d \sim 100$. In a relativistic outflow, due to relativistic aberration the photons move radially outwards. Therefore, the number of scatterings the up-scattered photons at an optical depth of $\tau$ undergoes is limited to $\tau$ (instead of $\tau^2$ in case of a non-relativistic outflow) before they get decoupled at the photosphere. Thus, even at these large $\tau$ the spectrum does not undergo saturated Comptonization, enabling us to detect the seed thermal component in the spectrum. 
We also note that any variability that happens at a radius below the photosphere cannot be traced via the observed light curve because the variability gets smeared out by the numerous scatterings the photons undergo before escaping from the photosphere (also see \citealt{Ruffini_2013}). 

\subsection{Second emission episode}
\label{sp_inter}
The second emission episode is found to have a spectrum of cutoff power law with the low energy power law index consistent with that of slow cooling synchrotron emission. The emission is also found to possess high polarisation fraction $> 43\%$ at $1.5\, \sigma$ confidence level of one parameter of interest. The high polarisation fraction suggests that the magnetic field is ordered on the angular scale of $1/\Gamma$. According to the discussions presented in section \ref{jetcomp}, we find that the second episode of emission is likely to be synchrotron produced by the energetic electrons generated in the internal shocks in the optically thin region of the outflow, as they cool in the large scale ordered magnetic fields originating from the central engine \citep{Toma_etal_2009,Granot_2003,Mao_etal_2018,Gill_etal_2019}.  

\begin{figure}
	\includegraphics[width=\columnwidth]{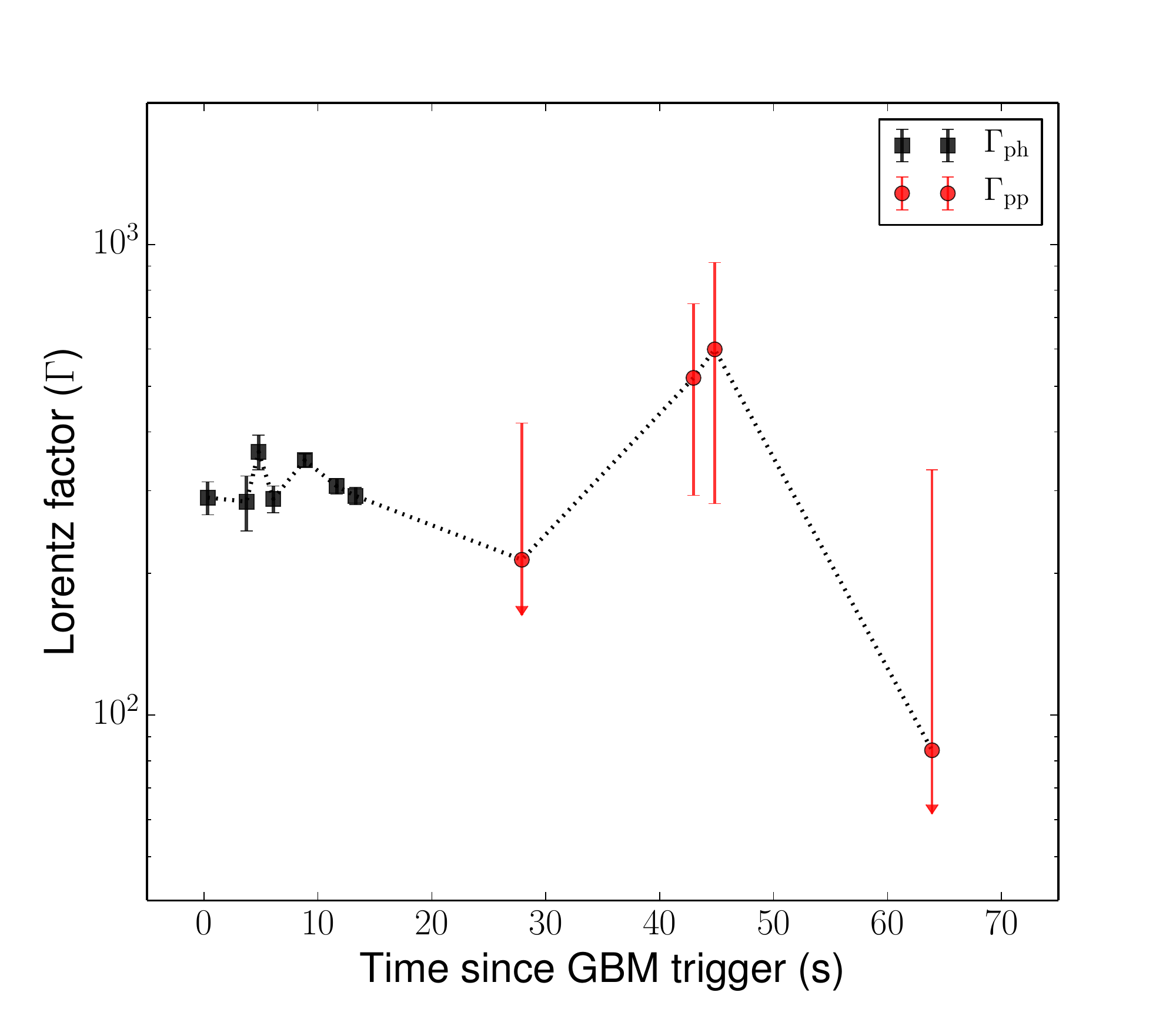}
    \caption{The temporal evolution of the Lorentz factor of the jet is shown. For the first episode, the Lorentz factor of the outflow estimated at the photosphere, $\Gamma_{ph}$ is shown with black squares. During the second episode, the Lorentz factor ($\Gamma_{pp}$) estimated using the pair opacity argument is shown in red circles.
    }
    \label{fig:gamma}
\end{figure}

The cutoff observed in the spectrum can be related to the intrinsic pair opacity created via photon-photon annihilation in the optically thin region at the dissipation site. We use the equation 2 given in  \cite{vianello2018bright} (or equation 126 in \citealt{granot2008opacity}), to estimate the Lorentz factor ($\Gamma_{pp}$) of the outflow and find it to be on average $354$ (Figure \ref{fig:gamma}). The parameter $C_2$ was assumed to be unity. Since the shocks are assumed to happen in the optically thin region, the variability of the outflow can be related to the variability timescale of the light curve. Using Bayesian Block$\;$algorithm on the {\it Fermi} GBM light curve $t_v$ is estimated to be $0.23\, \rm s$. This corresponds to a width of $c\, t_v \sim 10^{9}\, \rm cm$ across which the variation in the outflow happens. 
This results in internal shocks at a radius, $r_d \sim 2 \times 10^{15}\, \rm cm$ (Figure \ref{fig:radius}). 

The co-moving magnetic field intensity, $B'$, at the site of dissipation is estimated using the equation for the synchrotron peak energy, $E_{p,sync} \propto \Gamma \gamma_{el}^2 B'$, where $\gamma_{el}$ is assumed to be $\sim m_p/m_e$, where $m_p$ is the mass of the protron and $m_e$ is mass of the electron, as in internal shocks the electrons are expected to be mildly relativistic \citep{Daigne_Mochkovitch1998} and $\Gamma \sim 354$.
Thus, we find a co-moving magnetic field intensity of $B' \sim 7 \times 10^4$ Gauss.

We had initially assumed that the composition of the outflow is such that $\sigma_0 \ll 1$, and in the dissipation site the radiation is produced in the internal shocks. In order to check if our initial assumption is valid, we associate the estimated $B'$ to the magnetic field of the Poynting flux component that is possible at that radius. At such large dissipation radius of $10^{15}$ cm, the toroidal component of the magnetic field would be dominant ($B' \propto r^{-1}$). Thus, we estimate the initial lab frame magnetic field intensity, $B_0$ at the nozzle radius of the jet, $r_0$, to be $\sim 10^{15}$ Gauss which corresponds to a Poynting flux energy ($B_0^2/(8\, \pi) \times  4/3 \, \pi \, r_0^3$) of $\sim 2 \times 10^{51}\, \rm erg$. Since we consider the observed $E_{\gamma} \sim 2 \times 10^{53}\, \rm erg$ to be the fireball luminosity, we get an estimate of $\sigma_0 \sim 0.01$.  This is consistent with our initial assumption that $\sigma_0 \ll 1$.

Within the classical baryon dominated fireball model, photospheric emission is inherently expected in the observed spectrum. Below we discuss as to why we are not able to constrain the photospheric emission in the observed spectrum using the above inferred properties of the jet. The photospheric radius is given by $r_{ph} = L_{\gamma}/(8\pi m_p c^3 \Gamma^3 \kappa)$ where $L_{\gamma}$ is the observed luminosity of the second episode of emission. This yields $r_{ph} \sim 9 \times 10^{12}\, \rm cm$. If we assume that the nozzle radius of the jet remains the same as that estimated in the first episode of emission by being related mainly to the size of the central engine, we expect the saturation radius to be the same as that found in the first episode ($r_s \sim 3\times 10^{9}\, \rm cm$). 

The significance of the detection of thermal emission in the observed spectrum strongly depends on the adiabatic losses the thermal emission undergoes beyond the saturation radius before it gets decoupled at the photosphere, as well as on the photon statistics. The adiabatic factor is given by $(r_{ph}/r_s)^{-2/3} = \kappa \, F_{T}/F_{NT}$ 
\citep{ryde2004cooling, ryde2005thermal, iyyani2013variable}. Using the above estimates of photospheric and saturation radii, we estimate the expected $F_{T}/F_{NT}$ to be $\sim 5\%$ which is lower than that observed in first emission episode. The expected observable temperature of the thermal emission is estimated using the equation  
\begin{equation}
T_{ob}(r_{ph}) = \frac{1.48}{1+z}\left(\frac{L_{\gamma}}{\kappa \, 16\, \pi\, r_0^2 \sigma}\right)^{1/4}  \, \left(\frac{r_{ph}}{r_{s}}\right)^{-2/3}   
\end{equation}
where $\sigma$ is Stefan-Boltzmann constant \citep{pe2007new}. $T_{ob}$ is found to be $\sim 9 \, \rm keV$, comparable to the temperature observed in first emission episode. Thus, we suggest that the expected low observed temperature, relatively lower intrinsic flux of the photospheric emission with respect to the non-thermal emission and most critically the low photon statistics restrict us from finding any evidence for an additional thermal component in the observed spectrum of the second emission episode \citep{Ackermann2013_LATcatalog}.   

\section{Summary and Conclusions}
We have presented a detailed spectral and polarimetric analysis of the prompt emission and also a detailed study of the afterglow emission observed in X-rays of the GRB 160325A.  
The burst consisted of two main emission episodes separated by $9 \, s$ of mild activity or quiescence of the central engine.
The temporal evolution of the afterglow emission and the observed jet break affirmed that the jet was pointed towards the observer. This confirmation of the observing geometry played a crucial role in further deciphering the possible radiation process which led to the observed emission episodes. By composite modelling of the spectral and polarisation measurements of the two emission episodes as well as the afterglow observations, we arrive at the following inferences: (i) the jet is baryon dominated with a subdominant Poynting flux component; (ii) First emission episode originated at the photosphere. The observed emission is due to quasi-thermal Comptonisation happening in the dissipation site at moderate optical depths below the photosphere and thereby producing low polarisation; (iii) the second emission episode originated in the optically thin region above the photosphere and is due to synchrotron emission produced in internal shocks wherein the subdominant Poynting flux component provided a net ordered magnetic field configuration which resulted in high polarisation. 

The inferred dynamics of the jet shows a possible variation of the outflow at a lateral scale of $r_0 \, \theta_j$ during the first episode leading to the formation of shocks below the photosphere. On the other hand, during the second episode, we infer the variation in the outflow to happen on a larger radial scale than the nozzle radius of the jet, causing the internal shocks to form at a radius much above the photosphere.  
Thus, we have demonstrated that a composite physical model of spectral, polarimetric and afterglow observations, provide a clearer picture of the jet composition, radiation process as well as the region in the outflow from where the observed emission is produced.

\section*{Acknowledgements}
We would like to thank Prof. A. R. Rao and Dr Damien B\'egu\'e for the insightful discussions and comments on the manuscript. We would also like to extend our thanks to Dr Sunil Chandra for his help in analysis of UVOT data.  
This publication uses data from the {\it AstroSat} mission of the Indian Space Research Organisation (ISRO), archived at the Indian Space Science Data Centre (ISSDC). CZT-Imager is built by a consortium of institutes across India, including the Tata Institute of Fundamental Research (TIFR), Mumbai, the Vikram Sarabhai Space Centre, Thiruvananthapuram, ISRO Satellite Centre (ISAC), Bengaluru, Inter University Centre for Astronomy and Astrophysics, Pune, Physical Research Laboratory, Ahmedabad, Space Application Centre, Ahmedabad. 
This research has made use of {\it Fermi} data obtained through High Energy Astrophysics Science Archive Research Center Online Service, provided by the NASA/Goddard Space Flight Center. The Geant4 simulations for this paper were performed using the HPC resources at The Inter-University Centre for Astronomy and Astrophysics (IUCAA). This work utilized various software such as Python, astropy, corner, numpy, scipy, matplotlib, IDL, FTOOLS etc.
\addcontentsline{toc}{section}{Acknowledgements}





\bibliographystyle{mnras}
\bibliography{bib_GRB160325A} 



\appendix

\section{Spectral Models}
Here, we report the expressions of the photon flux of the different spectral models that were used in the paper in the units of $photons/cm^{2} /s/keV$.

\subsection*{(A) Black-body (bbody) model}
\label{sec:maths} 
\begin{equation}
    N_{BB}(E)=\frac{A \; E^{2}}{exp[E/kT]-1}
	\label{eq:bbody}
\end{equation}
where, A is the normalization, E is energy in keV, T is the temperature and k is Boltzmann constant.

\subsection*{(B) Cutoffpl model, a powerlaw high energy exponential rolloff energy}
\begin{equation}
    N_{C}(E) = A \; E^{-\alpha} exp[-E/\beta]
	\label{eq:cutoffpl}
\end{equation}
where, A is the normalization constant, $\alpha$ is the power law photon index and $\beta$ is the e-folding energy of exponential rolloff (in keV).

\subsection*{(C) Traditional Band (grbm) model}

\begin{equation}
    N_{B}(E) =
\begin{cases}
 A \, (E/100)^{\alpha} \, \rm exp[-\it E/E_{0}], \quad  & E < E_{0} \, (\alpha - \beta) \\ 
 A \, [(\alpha - \beta) E_{0}/100]^{\alpha - \beta} (E/100)^{\beta}  {\rm exp}[-(\alpha -\beta)], \quad  &  E >  E_{0} \, (\alpha - \beta)
\end{cases}
	\label{eq:band}
\end{equation}

where $\alpha$ is the low energy power law index, $\beta$ is the high energy power law index, $E_{0}$ is the break energy and $A$ is the normalisation. The spectral peak, $E_p = (2+\alpha) \; E_0$. \\

\section{Spectral analysis in 3ML software}
We conducted a crosscheck of the spectral analysis in 3ML \citep{Vianello_etal_2015} software, wherein the likelihoods for {\it Swift} BAT and {\it Fermi} GBM data are accounted according to the respective instruments' methodology to analyse their data. We obtained spectral fit results which were consistent with that obtained in the analysis done in Xspec (section 3). In Table \ref{tab:AIC_BIC_stats}, we report the AIC and BIC values obtained for the different spectral model fits. We note that $\Delta$ AIC and $\Delta$ BIC suggest that the best fit model in the brightest time interval of first and second emission episodes are blackbody + Cutoff power law and Cutoff power law alone respectively. This is in agreement to the reduced chi-square statistics that have been reported in section 3.  

\begin{table*}
\begin{small}
\caption{AIC and BIC statistics for each time interval}
\label{tab:AIC_BIC_stats}
\begin{center}
\begin{tabular}{p{1.25cm}ccccccc}
\hline \hline
Bin $\#$	& Time Interval (s)		& $AIC_{c}$		& $BIC_{c}$		& $AIC_{bb,c}$   & $BIC_{bb,c}$       & $\Delta$AIC, $\Delta$BIC
\\
            &                       & \tiny{Cutoffpl} & \tiny{Cutoffpl} & \tiny{Blackbody+Cutoffpl} & \tiny{Blackbody+Cutoffpl} 
\\ \hline 
1	& -2.54 - 3.21	& 3273 	& 3285  & 3273  & 3293  &  0,-8
\\ 
2	& 3.21 - 4.27	& 1467	& 1479   & 1468  & 1488  &   -1,-9
\\
3	& 4.27 - 5.34	& 1581	& 1593  & 1584 & 1604  &   -3,-11
\\ 
4	& 5.34 - 6.83	& 1827	& 1839   & 1827  & 1847  &  0,-8 
\\ 
5	& 6.83 - 10.87	& 2976	& 2988  & 2961  & 2981  &   15,7
\\ 
6	& 10.87 - 12.42	& 1938	& 1950   & 1933  & 1953  &   5,-3
\\
7	& 12.42 - 14.19	& 2055	& 2067  & 2048  & 2068  &   7,-1
\\ 
8	& 14.19 - 41.62	& 4767 	& 4778  & 4762  & 4782  &   5,-4
\\ 
9	& 41.62 - 44.32	& 2444	& 2456  & 2441  & 2461  &   3,-5
\\
10	& 44.32 - 45.35	& 1449	& 1461  & 1448  & 1468  &   1,-7
\\
11	& 45.35 - 82.46	& 5165	& 5177  & 5162  & 5182  &   3,-5
\\ \hline \hline
\end{tabular}
\begin{tablenotes}
      \small
      \item Note: $\Delta$ AIC = AIC$_{c}$ - AIC$_{bb,c}$; $\Delta$ BIC = BIC$_{c}$ - BIC$_{bb,c}$
\end{tablenotes}
\end{center}
\end{small}
\end{table*}

\section{polarisation During First Episode}
\begin{figure}
	\includegraphics[width=\columnwidth]{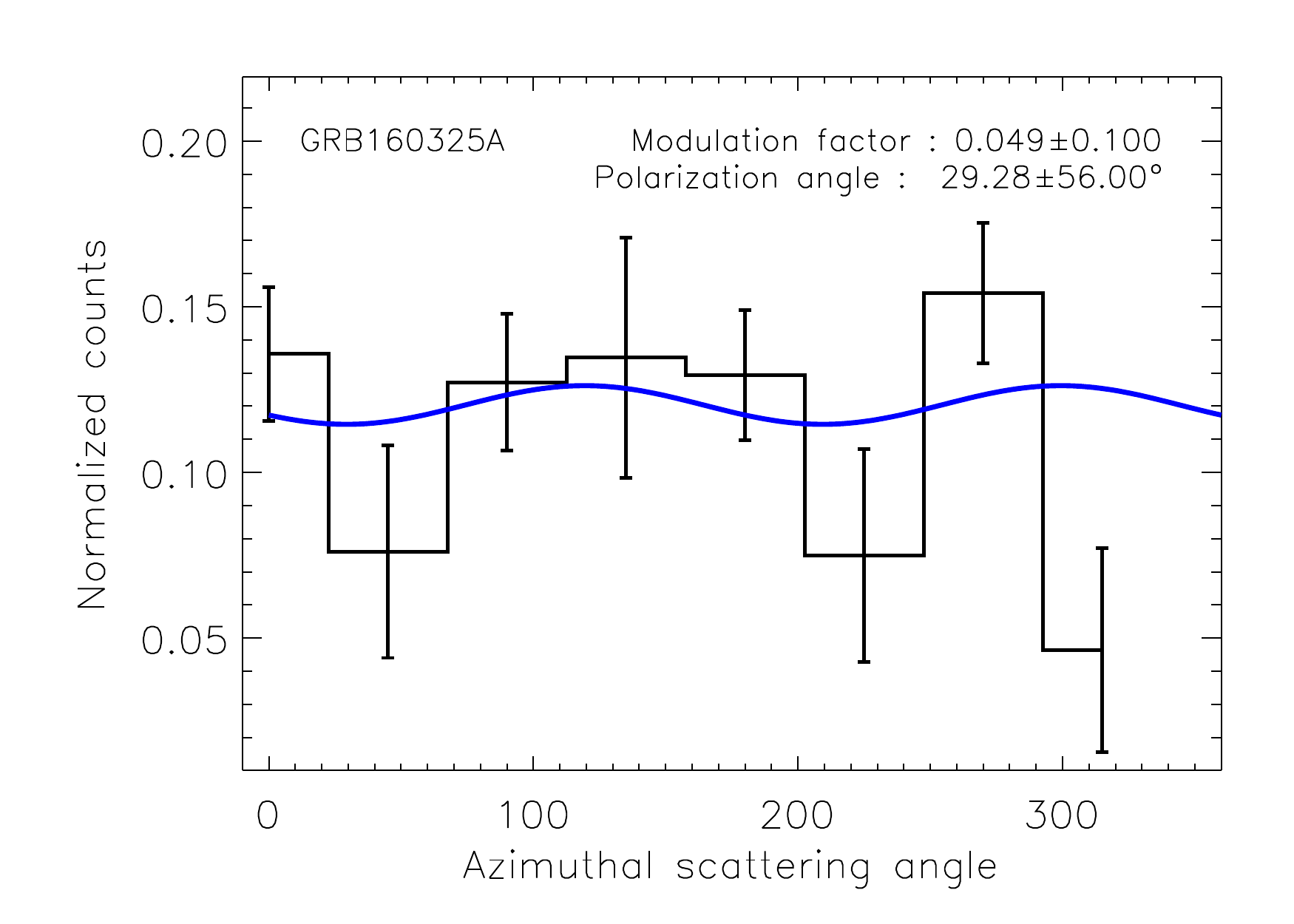}
    \caption{Azimuthal scattering angle distribution obtained for the first episode is shown. The best fit of the sinusoidal curve to the data is shown in blue solid line.}
    \label{fig:first_pulse}
\end{figure}

\begin{figure}
	\includegraphics[width=\columnwidth]{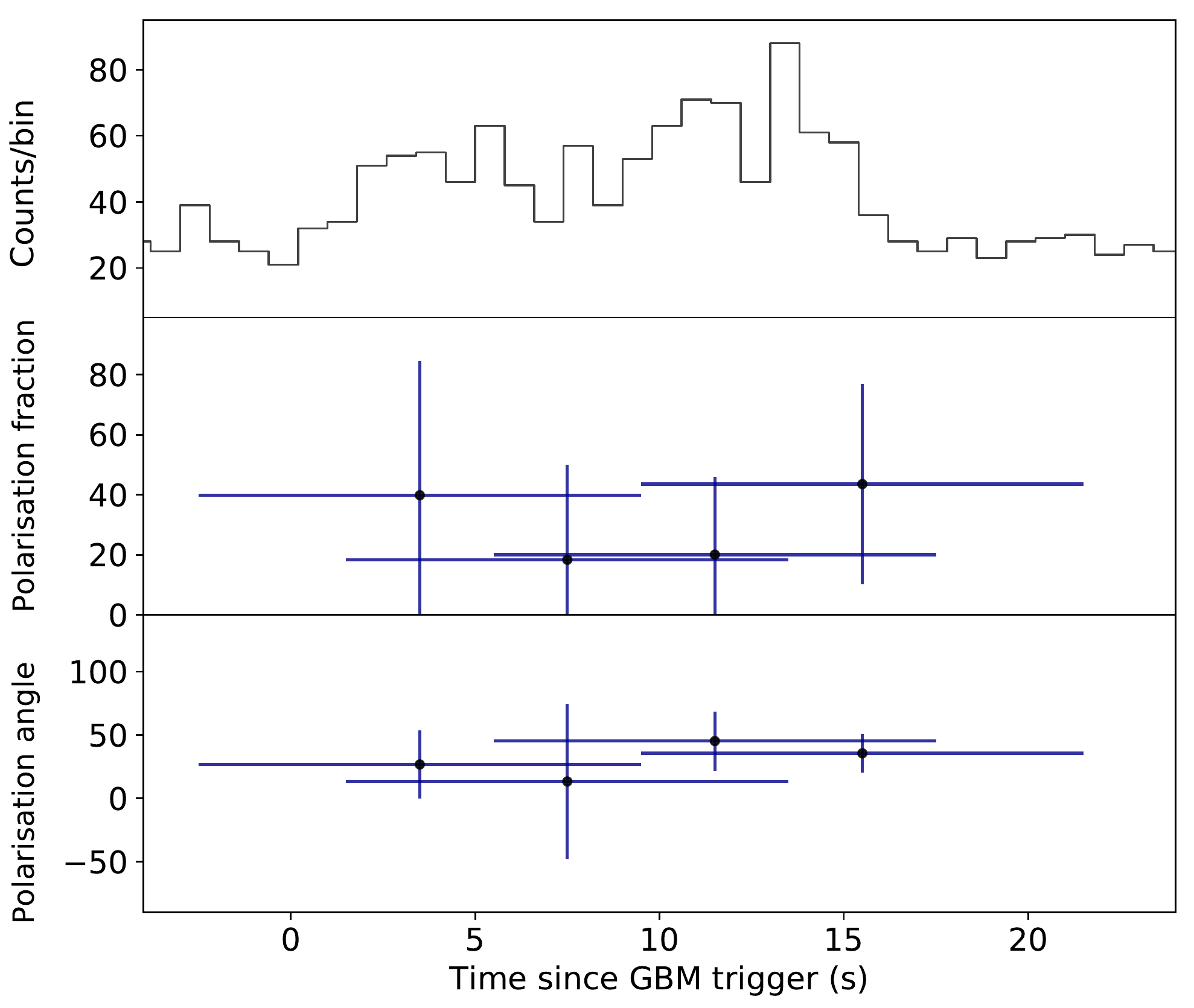}
    \caption{The Compton events lightcurve of the first episode with polarisation fraction and polarisation angle values obtained in the 12 s intervals such that each interval is shifted by 4 s are shown from top to down.}
    \label{fig:first_pulse_pa_pf}
\end{figure}

We do not observe a clear sinusoidal modulation in the azimuthal distribution of Compton events, and polarisation angle is also poorly constrained (Figure~\ref{fig:first_pulse}). The best fit parameter values are $\mu = 0.049 \pm 0.1$ and PA (detector plane) $ = 29.28 \pm 56.0^{\circ}$. The estimation of the Bayes factor for the polarised emission (sinusoidal model) and unpolarised emission (constant model) with the thermodynamic integration method in MCMC parameter space yields a value of 0.66. Hence, the Bayes factor favours unpolarised emission model for the first episode. A finer time-resolved polarisation analysis of first episode was also performed to check whether the unpolarised behaviour is intrinsic or due to temporal changes in the polarisation angle despite possessing high polarisation fraction. For this the polarisation fraction and the polarisation angle were measured for many 12 s intervals each shifted progressively by $4$ s, shown in the second and third panel from the top of Figure~\ref{fig:first_pulse_pa_pf}, respectively. The polarisation fraction indicates low values even in the finer time bins which in turn makes the polarisation angle highly uncertain. Hence, the first emission episode of the burst is intrinsically weakly polarised.

\section{Systematic Error on estimated Polarisation Fraction}
The possible sources of systematic errors on the polarisation fraction taken into account on the final reported value are as follows:(a) Due to multiple instances of photons interactions with the surrounding satellite structure (for on-axis case -CZTI collimators and coded mask) give rise to some uncertainty on the modulation. This uncertainty is found to be $\sim15\%$ for faint GRBs with $\sim400$ Compton events \citep{2019ApJ_Chattopadhyay}. (b) Other systematic errors involved are background selection, uncertainty on $\alpha$ of GRB spectrum, localisation and unequal quantum efficiency of CZTI pixels, which give rise to $\le 2\%$ uncertainty. (c) The uncertainties related to the spectrum and localisation will also be reflected in the $\mu_{100}$ modulations obtained in Geant4 simulations and is found to be $\le 1\%$ as the simulations are done for large number ($10^8$) of incoming photons. These uncertainties are written in detail in \citealt{2019ApJ_Chattopadhyay} and \citealt{2019ApJ_sharma}.
Considering all the above-mentioned uncertainties on $\mu$ and $\mu_{100}$, we find the propagated systematic error on polarisation fraction to be $\sim 11\%$. This thus brings the lower limit of PF quoted in the second interval to be $>43\%$ and $>20\%$ at $1.5 \, \sigma$ and $2 \, \sigma$ respectively for two parameters of interest. 
\bsp	
\label{lastpage}
\end{document}